\documentclass[a4paper, amsfonts, amssymb, amsmath, reprint, showkeys, nofootinbib, twoside,aps, prapplied]{revtex4-2}
\usepackage[dvipsnames,table,xcdraw]{xcolor} 
\usepackage[english]{babel}
\usepackage[utf8]{inputenc}
\usepackage[colorinlistoftodos, color=green!40, prependcaption]{todonotes}
\usepackage{amsthm}
\usepackage{mathtools}
\usepackage{physics}
\usepackage{xcolor}
\usepackage{graphicx}
\usepackage[left=23mm,right=13mm,top=35mm,columnsep=15pt]{geometry} 
\usepackage{adjustbox}
\usepackage{placeins}
\usepackage[T1]{fontenc}

\DeclareGraphicsExtensions{.pdf,.png,.jpg,.eps}
\usepackage{csquotes}
\usepackage{hyperref} 
\hypersetup{
	colorlinks=true,
	linkcolor=blue,
	citecolor=blue,
	filecolor=blue,
	urlcolor=blue,
}
  
\usepackage{color}
\usepackage{ulem}

\newcommand\old{\bgroup\markoverwith{\textcolor{ForestGreen}{\rule[0.5ex]{2pt}{0.8pt}}}\ULon}

\begin{document}

\title{{In situ subwavelength microscopy of ultracold atoms using dressed excited states}} 
	
	\author{R. Veyron}
	\author{J-B. Gerent}
\author{G. Baclet}
\author{V. Mancois}
\author{P. Bouyer}
\author{S. Bernon}
\affiliation{ LP2N, Laboratoire Photonique, Num\'{e}rique et Nanosciences, Universit\'{e} de Bordeaux-IOGS-CNRS:UMR 5298, rue F. Mitterrand, F-33400 Talence, France}
	
	\begin{abstract}
In this work, we implement a new method for imaging ultracold atoms with subwavelength resolution capabilities and determine its regime of validity. It uses the laser driven interaction between excited states to engineer hyperfine ground state population transfer in a three-level system on scales much smaller than the optical resolution. Subwavelength imaging of a quantum gas is atypical in the sense that the measurement itself perturbs the dynamics of the system. To avoid induced dynamics affecting the measurement, one usually measures "rapidly" the wavefunction in a so-called strong imaging regime. We experimentally illustrate this regime using a thermal gas ensemble, and demonstrate subwavelength resolution in quantitative agreement with a fully analytical model. Additionally, we show that, counter-intuitively, the opposite weak imaging regime can also be exploited to reach subwavelength resolution. As a proof of concept, we demonstrate that this regime is a robust solution to select and spatially resolve a 30 nm wide wavefunction, which was created and singled out from a tightly confined 1D optical lattice. Using a general dissipation-included formalism, we derive validity criteria for both regimes. The formalism is applicable to other subwavelength methods. 

\end{abstract}
\date{\today}
	\maketitle
	
	\section{Introduction}
	Quantum gas microscopes have emerged as essential tools for quantum simulations using cold neutral atoms in optical lattices \cite{Schaefer2020}. For instance, anti-ferromagnetic long-range order has been measured in standard optical lattices \cite{Mazurenko2017,Koepsell2020} by measuring spin-correlations between lattice sites with a microscopy that can resolve both atomic density and spin. In this context, subwavelength lattices using for instance stroboscopic techniques \cite{Nascimbene2015,Tsui2020}, {Raman-coupled multi-level states \cite{Anderson2020}} or near-field lattices in front of a surface \cite{Bellouvet2018} have emerged as a possibility to enhance interactions which is essential to enter into strongly correlated phases. 
	Imaging such systems thus requires developing novel techniques to beat the far-field diffraction limit of $\lambda/2$ where $\lambda$ is the imaging wavelength. The field of bio-imaging has long since been confronted with this issue and has developed dedicated methods \cite{Godin2014} to bypass this limit {like structured illumination microscopy \cite{Gustafsson00}, stimulated emission depletion \cite{Hell94}, or single-molecule localization \cite{Rust2006}.}
	
	{Among the methods applied to cold atoms, some take advantage of a tightly focused beam to image individual sites \cite{Weitenberg10}, others use the optical transfer function noise properties and discreteness of the object to gain in resolution \cite{Alberti2016} or magnify the object using matter wave optics \cite{Asteria2021}. We often refer to the works \cite{McDonald2019,Subhankar2019} which have {experimentally} pioneered superresolution imaging of quantum gas. Exploiting the non-linearity of light-matter interaction, these methods locally transfer atoms from a dark to a bright state and have demonstrated resolutions down to tens of nanometers. In \cite{McDonald2019}, a strong standing wave of optical pumping light incoherently transfers all atoms except subwavelength slices of atoms that remain in the dark state. In \cite{Subhankar2019}, a two-photon dark state using a standing wave coherently populates the bright state in subwavelength slices. }{In these methods, the imaging position is determined by the position of the standing waves minima.}
	
	{In this work, we both demonstrate and characterize a novel subwavelength imaging method and present a general theoretical formalism based on a Schr\"{o}dinger equation with dissipation to describe the system dynamics. The formalism allows us to show that subwavelength control can be reached both in the strong and weak imaging limits, which respectively correspond to a diabatic and adiabatic evolution of the system. While only the first has been studied in \cite{McDonald2019,Subhankar2019}, we show that the second achieves similar performances with reduced timing constraints. The formalism additionally  allows us to define the validity ranges of the two imaging regimes. The novel demonstrated method relies on an incoherent transfer of  subwavelength slices of atoms from a dark to a bright state. The method is based on dressed state engineering which is well adapted to multi-level systems and that has been theoretically investigated for near-field traps \cite{Bellouvet2018}: a strong light shift is generated by a spatially varying profile of optical intensity {whose} radiation frequency is tuned close to an atomic transition between two excited states of $^{87}$Rb at 1529 nm.
	First, in section \ref{sec:methods}, we present the dressing method, modeled by a three-level system and provide a theoretical formalism that describes the system dynamics during imaging. Validity criteria for the two opposite imaging strength regimes are derived as well as their expected performance. Experimental results are presented in section \ref{sec:results}. In \ref{subsec:apparatus}, we present the main characteristics of the cold atom apparatus described in more detail in \cite{Veyron2022prr}. Then, in section \ref{sec:ho_cloud}, in the strong imaging regime, we measure the spatial resolution as a function of the light shift gradient and demonstrate that the resolution can easily be tuned well below the diffraction limit in good agreement with the model. Finally, in section \ref{sec:1Dlatt_cloud}, in the weak imaging regime, we prepare and test the subwavelength resolution on the narrowest atomic density we could prepare via the adiabatic loading of a Bose-Einstein Condensate in the first band of a tightly confined lattice.}

\section{Method}\label{sec:methods}
\begin{figure}
    \begin{center}
        \includegraphics[scale=0.34]{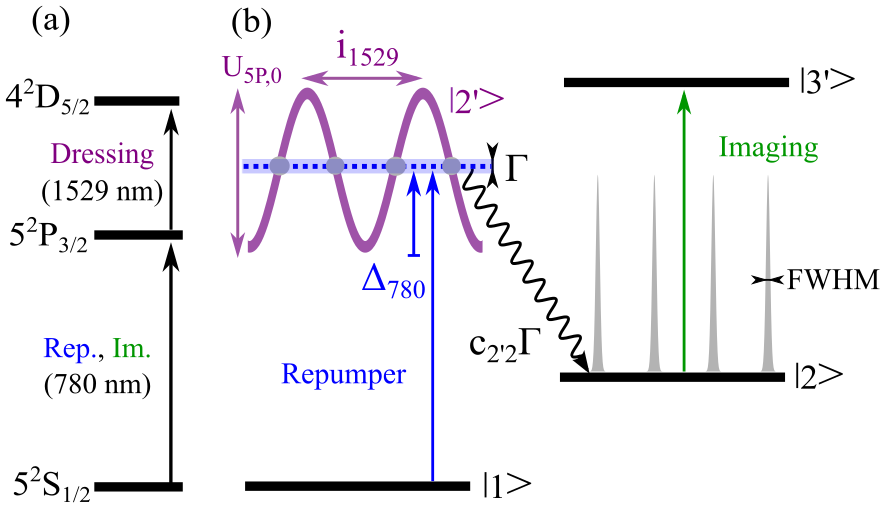}
        \caption{(a) Three fine structure states of $^{87}$Rb: the ground state $5^2S_{1/2}$ and the two excited states $5^2P_{3/2}$ and $4^2D_{5/2}$. (b) A three-level system with an optically dressed excited state using a 1529 nm lattice probed with a repumper at a detuning $\Delta_{780}$ and imaged on a cycling transition.}
        \label{fig:3levelAtomDiagram_5levelAtomDiagram}
    \end{center}
\end{figure}

The method is depicted on Fig. \ref{fig:3levelAtomDiagram_5levelAtomDiagram}. It consists in a spatially-dependent incoherent transfer of atomic populations between the hyperfine ground states $\ket{1}$ and $\ket{2}$ via their common coupling to a dressed excited state $\ket{2'}$. The energy of $\ket{2'}$ is spatially modulated such that a homogeneous repumper excitation is only locally resonant and the population transfer occurs on scales that are not limited by diffraction. After removing the dressing field, the  population transferred in $\ket{2}$ is imaged on a cycling transition from $\ket{2}$ to $\ket{3'}$.

Experimentally, as shown on Fig. \ref{fig:3levelAtomDiagram_5levelAtomDiagram}a, a 1529 nm laser lattice intensity profile drives the transition between the two excited states $5^2P_{3/2}$ and $4^2D_{5/2}$ of $^{87}$Rb, resulting in a spatially-dependent light shift for the excited state $\ket{2'}$ (Fig. \ref{fig:3levelAtomDiagram_5levelAtomDiagram}b). The excited state potential is therefore given by: 
\begin{equation}
    \begin{aligned}
        U_{\text{5P}}(x) = & U_{\text{5P,0}} \cos^2 \left( \frac{k_{1529}}{2}x -\frac{\pi}{4} \right),
        \label{eq:LightShiftCoPropa}
    \end{aligned}
\end{equation}
where  $k_{1529}={2\pi}/{i_{1529}}$ is the lattice wave vector, $i_{1529}$ is the lattice period, and $U_{\text{5P,0}}$ the total amplitude which is given by the maximum of the 1529 nm laser intensity profile $I_{1529}$. This amplitude is computed from the diagonalization of the standard electric-dipole and hyperfine Hamiltonians as detailed in Appendix \ref{sec:light_shifts}. We restrict our model to the linear light shift regime where the amplitude is proportional to the intensity.

The spatially-dependent detuning of the 780 nm laser with respect to the bare hyperfine states is:
\begin{equation}
    \begin{aligned}
        \Delta(x)=\Delta_{780} - \Delta_{\text{LS}}(x),
        \label{eq:LightShiftCoPropa_detuning}
    \end{aligned}
\end{equation}
where $\Delta_{780}=\omega_{780}-\omega_{0}$ is the bare detuning between the laser  frequency $\omega_{780}$ and the bare atomic transition frequency $\omega_{0}$, $\Delta_{\text{LS}}(x)=U_{\text{5P}}(x)-U_{\text{5S}}(x)$ is the differential light shift between the ground and excited states, and $U_{5S}(x)$ is the potential of the ground state which is small compared to the excited state light shift. {In the low saturation limit ($s_0\ll1$), the spatially-dependent saturation parameter is given by:
\begin{equation}
    \begin{aligned}
        s(x)=&\frac{s_{0}}{1+\left(\frac{2\Delta(x)}{\Gamma}\right)^{2}} \approx \frac{s_{0}}{1+\left(\frac{x}{X_0}\right)^2},
        \label{eq:LightShiftCoPropa_saturation}
    \end{aligned}
\end{equation}
where $\Gamma/2\pi=6.066$ MHz and $s_0=I_{780}/I_{\text{sat,rep}}$ are respectively the natural linewidth and the on-resonance saturation parameter for the repumper transition, with $I_{780}$ the intensity of the repumper beam and $I_{\text{sat,rep}}$ the saturation intensity of the optical transition.  
The last expression in Eq. (\ref{eq:LightShiftCoPropa_saturation}) will be used in the rest of the manuscript. It corresponds to a repumper laser tuned at the middle of the excited state modulation such that $\Delta_{780}=U_{\rm{5P,0}}/2$ where the spatial selectivity of the method is maximal. Around $x=0$ the detuning can be  linearised ($\Delta(x) = \pi U_{\rm{5P},0} x/i_{1529}$) and a transfer occurs on a scale given by the localization length: 
\begin{equation}
    \begin{aligned}
        X_0 = \frac{i_{1529} \Gamma}{2 \pi U_{\rm{5P},0}}. 
        \label{eq:localisation_parameter}
    \end{aligned}
\end{equation}
We now focus on deriving the point spread function (PSF) of this sub-wavelength imaging method. In ultra-cold atomic physics the particles' wavefunctions are larger than achievable subwavelength resolutions which is a different regime than encountered in bio-imaging. Such localized modification of the wavefunction leads to a spatio-temporal dynamics of the wavefunction during the imaging process that one needs to account for. To simplify the study, \cite{McDonald2019,Subhankar2019} have tried to decouple the spatial and temporal dynamics using the strong imaging regime that they reach via a rapid transfer. We derive below the validity criteria of such regime and show that rapidity is not the sole criteria to reach subwavelength imaging.\\}

\paragraph{Temporal transfer dynamics :}
{In Appendix \ref{sec:Rb87_D2_3LS}, we give the multi-level structure of $^{87}$Rb and justify the choice of a three-level system (3LS) model to describe the population transfer from $\ket{1}$ to $\ket{2}$. As detailed in Appendix \ref{subsec:optical_Bloch_eq_dynamics} and in \cite{McDonald2019}, for all parameters used in this work, the transfer rate of population at position $x$ is independent of time  and can be very well approximated by:
\begin{equation}
    \begin{aligned}
        \kappa({x}) & = \frac{c_{2'2}\Gamma}{2} s({x}) \approx \frac{\kappa_0}{1+\left(\frac{{x}}{X_0}\right)^2},  
        \label{eq:dissipation_equation_main}
    \end{aligned}
\end{equation}
where the on-resonance transfer rate is $\kappa_0=c_{2'2}\Gamma s_{0}/2$.
Eq. (\ref{eq:dissipation_equation_main}) leads to an exponential growth of the population in $\ket{2}$ with a rate proportional to the branching ratio ($c_{2'2}$) between $\ket{2'}$ and $\ket{2}$ and the scattering rate at position $x$.}\\ 

\paragraph{Spatio-temporal dynamics :}
{In the following, we only consider the spatial dynamics of $\ket{1}$. Indeed, once atoms have been transferred to $\ket{2}$, their spatial evolution will not impact the imaging resolution which is only sensitive to the number of transferred particles. Following previous works \cite{Keeling2008, Savenko2013, Sierra2015}, the effect of the transfer rate $\kappa(\hat{x})$ on the evolution of the wavefunction $\Psi(x,t)$ for atoms in $\ket{1}$ can be accounted for by a loss term in the Schrodinger equation:
\begin{equation}
    \begin{aligned}
        i\hbar\frac{d\ket{\Psi(x,t)}}{dt} = \hat{H}_{\text{0}}\ket{\Psi(x,t)} - i\frac{\hbar\kappa(\hat{x})}{2}\ket{\Psi(x,t)}
        \label{eq:SchrodingerEquation_Main}
    \end{aligned}
\end{equation}
where $\hat{H}_{\text{0}}$ is the Hamiltonian including kinetic and potential energies of a harmonic oscillator, and {$\kappa(\hat{x})$ is a rate operator derived from Eq. (\ref{eq:dissipation_equation_main})}. $\hat{H}_{\text{0}}$ satisfies $\hat{H}_{\text{0}}\ket{\phi_n}=E_n\ket{\phi_n}$ where $\ket{\phi_n}$ are eigenstates, and $E_n=\hbar\omega_{\text{HO}}(n+1/2)$ are the eigenvalues with $\omega_{\text{HO}}$  the harmonic oscillator trap frequency.} \\

{This formalism allows to define the strong and weak imaging regimes:
\begin{itemize}
    \item strong imaging: a high transfer rate $\kappa_0$ is used during a very short imaging time. The wavefunction is strongly affected but has no time to evolve before the end of the transfer.
    \item weak imaging: a weak transfer rate $\kappa_0$  is applied during a long imaging time, allowing the wavefunction to remain in the ground state during all the process.
\end{itemize}
Their applicability domain is presented below and we show that both are relevant to perform subwavelength imaging.}

\subsection{Strong imaging regime}\label{txt:MethodStrong}
{As discussed in \cite{McDonald2019,Subhankar2019}, a rapid transfer can be used to super-resolve a wavefunction. This process creates a dip of width $\Delta x$ in the initial wavefunction that corresponds to a velocity spread $\Delta v$ that is minimal for a Gaussian spatial localization. In this case, we achieve the equality of the Heisenberg uncertainty principle : $\Delta v  \Delta x = \hbar/(2m)$ where $m$ is the mass of the particle. The strong imaging regime ($t_{\rm tr} \Delta v\ll \Delta x$) is then fulfilled for transfer  time:
\begin{equation}
    \begin{aligned}
        t_{\rm tr} \ll 2 m\Delta x ^2/\hbar
        \label{eq:Heisenberg}
    \end{aligned}
\end{equation}
For a $10$ nm Gaussian standard deviation localization of $^{87}$Rb atoms, this amounts to 273 ns and is more restrictive for non-Gaussian localization.}

{The quite general case of a Lorentzian transfer rate of full-width at half maximum $\rm FWHM_s$, achieved here with optical pumping (Eq. (\ref{eq:dissipation_equation_main})), leads to a velocity spread $\Delta v$ derived in Appendix \ref{an:strongimaging}. In the strong imaging regime, the imaging time should satisfy $t_{\rm tr} \Delta v \ll \rm FWHM_s$ which yields the following condition on the transfer rate: 
\begin{equation}
    \begin{aligned}
        \mathcal{S}=\frac{2}{\ln(2)}\frac{\hbar\kappa_0}{\hbar^2/(2 m X_0^2)}\gg 1,
        \label{eq:Heisenberg2}
    \end{aligned}
\end{equation}
where we have shaped the equation to emphasize the energy ratio of pumping strength $\hbar \kappa_0$ over localization recoil $\hbar^2/(2m X_0^2)$.
One should note that this criteria does not restrict the imaging time but rather constrains the imaging strength $\kappa_0$, hence the name "strong", rather than "fast", imaging regime.}

{The strong imaging regime is experimentally demonstrated in section \ref{sec:ho_cloud}. The atomic density in $\ket{2}$ at position $x$ and time $t$, for a  repumper resonant at position $x_r$ is :
\begin{equation}
\begin{aligned}
\rho^{s}_{22}(x,x_r,t) =  \rho_{00}(x,t=0) (1 - e^{ -\kappa(x-x_r) t }),
\label{eq:3level_rhoee}
\end{aligned}
\end{equation}
where $\rho_{00}(x,t=0)$ is the initial atomic density for atoms in $\ket{1}$. For a homogeneous initial density ($\rho_{00}(x,t=0)$ constant), the width of $\rho_{22}$ as a function of $x$ is the characteristic scale of the super-resolution imaging.}

{For long pulse duration compared to the internal state dynamic ($s_0\Gamma t_{\rm tr} \gg 1$) and for well-resolved fringes such that $\Gamma/U_{\text{5P,0}}\ll1$, we can compute the full-width at half maximum of Eq. (\ref{eq:3level_rhoee}) at the middle of the modulation where $\Delta_{780}=U_{\text{5P,0}}/2$:
\begin{equation}
    \begin{aligned}
        \text{FWHM}_{\text{s}} & = 2 X_0 \sqrt{\frac{\kappa_0 t_{\rm tr}}{\ln(2)}}.
        \label{eq:3level_FWHM_limit_middle}
    \end{aligned}
\end{equation}
As detailed in section \ref{sec:ho_cloud}, for our experimental implementation, Eq.  (\ref{eq:3level_FWHM_limit_middle}) yields spatial widths smaller than the diffraction limit ($\lambda/2=390$ nm).}


\subsection{Weak imaging regime}
{In the weak imaging regime, the perturbation introduced by imaging occurs at a slow rate $\kappa$, preventing the transfer of population to any excited states of motion. In other words, this regime is achieved when the evolution of the wavefunction in the internal state $\ket{1}$ is adiabatic with respect to any of the $n^{\rm th}$ vibrational level of the harmonic oscillator $\ket{\phi_n}$  \cite{Dahan96}:
\begin{equation}
    \begin{aligned}
        | \bra{\phi_n}\frac{d}{dt}\ket{\Psi(x,t)}|=| \bra{\phi_n}\frac{\kappa(\hat{x})}{2}\ket{\phi_0}|\ll \frac{E_n-E_0}{\hbar}.
        \label{eq:Adiabatic_Main}
    \end{aligned}
\end{equation}}

{The equality in Eq. (\ref{eq:Adiabatic_Main}) is obtained using Eq. (\ref{eq:SchrodingerEquation_Main}), the harmonic oscillator modes orthogonality and a state initially prepared in the ground state $\ket{\Psi(x,0)}=\ket{\phi_0}$ with harmonic oscillator width $a_{\text{HO}}=\sqrt{\hbar/(m\omega_{\text{HO}})}$. This condition is analytically derived in Appendix \ref{subsec:calibration_piezo}, and for $r_0\ll1$ simplifies to:
\begin{equation}
    \begin{aligned}
        \mathcal{W}_2&=\sqrt{\frac{\pi}{8}} \frac{r_0\kappa_0}{2\omega_{\text{HO}}} \ll 1,\\
        \mathcal{W}_1&= \sqrt{\pi}e^{-\frac{1}{2}}\frac{r_0\kappa_0}{2\omega_{\text{HO}}} \ll 1,
        \label{eq:adiabaticity_criteria_n_is_2_largelimit_main}
    \end{aligned}
\end{equation}
where $r_0=X_0/a_{\text{HO}}$ is the ratio of the localization length over the wavefunction width. $r_0\ll1$ therefore corresponds to resolving the wavefunction details. $\mathcal{W}_2$ and $\mathcal{W}_1$ are the adiabaticity criteria respectively for centered and off-centered coupling terms. We emphasize once more that this regime depends on the imaging strength $\kappa_0$ and not on the transfer time $t_{\rm tr}$.}

{In this regime, the number of atoms $\rho_{22}(x_r,t)$  transferred to  $\ket{2}$ at time $t$ for a resonant repumper position $x_r$ is given by:
\begin{equation}
    \begin{aligned}
        \rho^{w}_{22}(x_r,t) &= N_{0} \left(1-e^{-\bra{\phi_0}\kappa(\hat{x}-x_r)\ket{\phi_0}t}\right)\\
        &\approx N_{0} \bra{\phi_0}\kappa(\hat{x}-x_r)\ket{\phi_0}t
        \label{eq:atom_number_weak_regime}
    \end{aligned}
\end{equation}
where $N_{0}$ is the atom number in state $\ket{1}$ in the ground state of the harmonic oscillator.}

{The rate is maximum for $x_r=0$: $\bra{\phi_0}\kappa(\hat{x})\ket{\phi_0}=\sqrt{\pi}r_0e^{r_0^2}\text{erfc}(r_0)\kappa_0$. The approximated linearized expression in Eq. (\ref{eq:atom_number_weak_regime}) is satisfied in the limit of weak depletion ($\bra{\phi_0}\kappa(\hat{x})\ket{\phi_0}t_{\rm tr}\ll1$). $\bra{\phi_0}\kappa(\hat{x}-x_r)\ket{\phi_0}$ corresponds to the Voigt function which size is given within $1.2\%$ error by \cite{WHITING1968}:
\begin{equation}
    \begin{aligned}
        \text{FWHM}_{\text{w}}^{\rm total} &= X_0+\sqrt{X_0^2+4a_{\rm HO}^2\ln(2)}\\
        &\approx 2a_{\rm HO}\sqrt{\ln(2)}+\left(1+\frac{r_0}{\sqrt{\ln(2)}}\right)X_0
        \label{eq:FWHMweak_main}
    \end{aligned}
\end{equation}
Eq. (\ref{eq:FWHMweak_main}) gives the typical width of the measured density profile, which is the initial harmonic oscillator width broadened by the imaging resolution length $X_0$.}

{We note that the validity criteria of weak imaging upper bounds the scattering rate by the harmonic oscillator frequency. This partly reduces the capability of the weak regime to measure fast enough to capture the oscillation dynamics in the lattice. A possible solution to reduce the imaging timescale is to use large atomic ensemble $N_0\gg1$. In that case, even at small $t_{\rm tr}$, some atoms get transferred by statistical process. Nevertheless, for short timescale imaging, the strong imaging regime is certainly to favour.}

\section{Results}\label{sec:results}

{The experimental setup used to produce and image the atomic clouds is presented in \ref{subsec:apparatus}. Atom numbers are precisely measured using a calibrated in situ absorption imaging \cite{Veyron2022prr}. The next two parts are demonstrations of the strong and weak imaging regimes. In \ref{sec:ho_cloud} (strong imaging), we consider the case of a cigar-shaped thermal atomic cloud and a 1529 nm lattice that can be resolved by our imaging system. To validate our model, we measure the number of atoms transferred into the state $\ket{2}$ and compare it to the expected atom number deduced from our model. Finally, in \ref{sec:1Dlatt_cloud} (weak imaging) we use the method to prepare and test the subwavelength resolution on the narrowest atomic density we could prepare via the adiabatic loading of a Bose-Einstein Condensate in the first band of a tightly confined 1D lattice with a spacing smaller than the diffraction limit.}
 


\subsection{Absolute calibration of the scattering cross section} \label{subsec:apparatus}
We prepare an atomic cloud of $^{87}$Rb in state $\ket{1}$ using a hybrid trap composed of a magnetic trap compensating the gravity field and a crossed dipole trap (DT1 and DT2) \cite{Lin2009}.  Evaporative cooling is performed to produce either a thermal cloud or a BEC. 

As shown on Figs. \ref{fig:atom_number_BAT_co_propa} and \ref{fig:counterPropasetup}, the 1529 nm lattice intensity is generated by the interferences of co- or counter-propagating laser beams with a linear polarization aligned along $z$ which is set as the quantization axis. A $\pi$-polarized repumper pulse transfers the atoms to the state $\ket{2}$ which is imaged using a cycling transition. The saturation parameters of the repumper and imaging beams are computed for each experimental run by monitoring their optical power.

We use a high numerical aperture absorption imaging system with a resolution limit of 1.3 \textmu m \cite{Li2018}. An image with atoms $I_{\text{at}}$, a reference image $I_{\text{noat}}$ of the imaging beam and an image for the background $I_{\text{back}}$ are acquired to compute the transmission $T=(I_{\text{at}}-I_{\text{back}})/(I_{\text{noat}}-I_{\text{back}})$ and the optical density (OD) given by $b=-\alpha \ln(T)+s_{\text{im}}(1-T)$ where the saturation parameter for the imaging beam is set at $s_{\text{im}}=1$ {for a duration of 8 \textmu s.}

For accurate atom number measurements, it has been crucial to calibrate the reduction factor $\alpha$ of the scattering cross section  which scales linearly with the optical density \cite{Veyron2022prr,Veyron2022}. In \cite{Veyron2022prr}, we measured $\alpha(b)=\alpha_0+\beta b$ where $\alpha_0=1.17$ and $\beta=0.255$. Using this correction, the optical density is reformulated as $b={(-\alpha_0 \ln(T)+s_{\text{im}}(1-T))}/{(1+\beta \ln(T))}$. Finally, the experimental atom number $N_{\rm at}=\iint b(x,y)dxdy/\sigma_0$ can be computed  for any region of the image and uses the scattering cross section of a $\sigma^-$-polarized probe ($\sigma_0=2.907\times10^{-9}$ cm$^2$).


\subsection{
{Strong imaging regime with thermal atoms}}\label{sec:ho_cloud}

For this experiment, we prepare and characterize by time-of-flight an initial thermal cloud containing $N_{\rm at}^{\rm tot}=1.03(16)\times10^{5}$ atoms in $\ket{1}$ at a temperature of 169(10) nK, just above the condensation threshold. The cloud is trapped and compressed solely in DT2, creating a cigar-shape elongated along $y$. The atomic density is then homogeneous over a few 1529 fringe periods ($i_{1529}=8.3$ \textmu m). A homogeneous magnetic bias of 280 mG along $z$ defines the quantization axis. The maximum density at the cloud center is $1.8 \times 10^{19}$ at/m$^3$, giving an optical density of 31. To avoid high OD distorsion of atom number counting, we reduce the OD using coherent micro-wave (MW) transfer between the states $\ket{1}$ and $\ket{2}$ with transfer probability function $P(t_{\text{MW}})=P_m\sin^2(\pi t_{\rm MW}/T_{\rm MW})$, where $T_{\rm MW}=56$ \textmu s is the period and $P_m=0.96$ is the maximum probability. A first MW $\pi$-pulse transfers all atoms from $\ket{1}$ to $\ket{2}$ and is followed by a short optical repumper pulse that empties the ground state $\ket{1}$. A second MW pulse of duration $t_{\text{MW}}=8$ \textmu s is used to transfer a controlled population back into $\ket{1}$ and a {resonant $\ket{2}$ to $\ket{3'}$ laser} pulse pushes away the remaining atoms in $\ket{2}$. This sequence reduces the maximum optical density down to 6. In this configuration, the measured in situ cloud widths are $\sigma_y=64$ \textmu m and $\sigma_x=2.4$ \textmu m. 
The subwavelength imaging method described in section \ref{txt:MethodStrong} is then performed on this thermal sample. {The use of the wavefunction model is justified as the de Broglie wavelength ($\lambda_{\rm dB}= 460 $nm) is higher than the localization length ($X_0=33$ nm for $U_{\text{5P,0}}=40\Gamma$). This also allows to neglect the residual thermal energy contribution to the spacio-temporal dynamics in Eq. (\ref{eq:SchrodingerEquation_Main}).} 

\begin{figure}
\begin{center}
\includegraphics[scale=0.39]{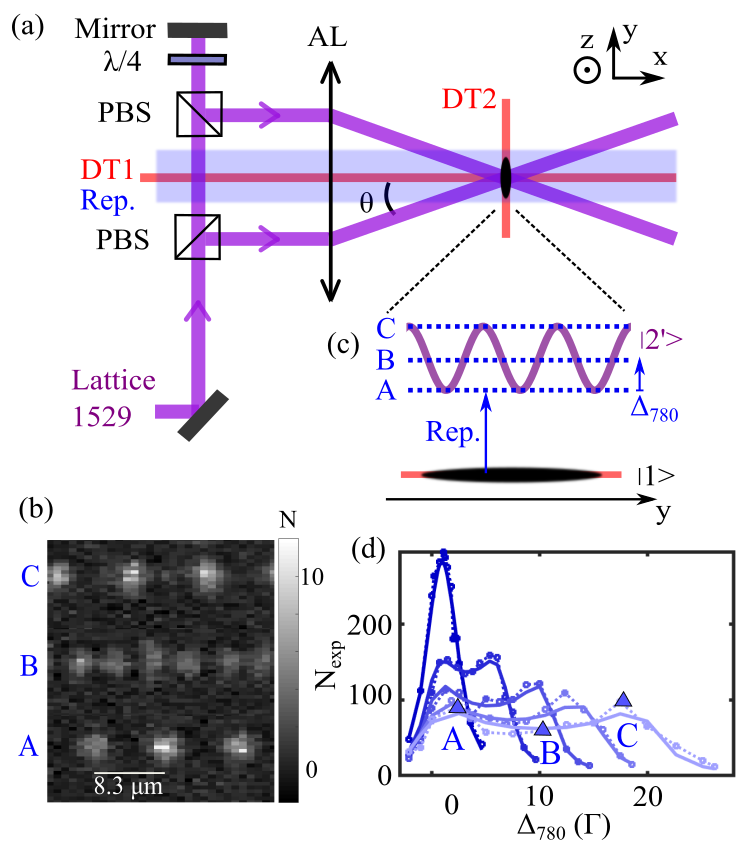}
\caption{(a) Optical setup for the generation of the 1529 nm lattice using a combination of mirrors, polarizing beam splitters (PBS), a quarter-wave plate ($\lambda/4$) and an aspherical lens (AL). (b) 2D absorption images of repumped atom numbers per pixel for a homogeneous atomic cloud. The letters A, B, and C correspond to the bottom, middle and top of the lattice as shown in (c) showing the different resonance conditions between the dressed excited state $\ket{2'}$ and the ground state $\ket{1}$. (d) Integrated atom number per lattice period (dashed lines with circles) for depths $U_{\text{5P,0}}$ of $2.5\Gamma$ (dark blue), $7.2\Gamma$, $11.6\Gamma$, $15.8\Gamma$ and $21\Gamma$ (light blue), and fits (solid lines) with Eq. (\ref{eq:OneFringeBloch}) where the free parameters are the light shift and the total atom number. The light blue curve ($21\Gamma$) corresponds to the data of panel (b).}
\label{fig:atom_number_BAT_co_propa}
\end{center}
\end{figure}

In the co-propagating case depicted in Fig. \ref{fig:atom_number_BAT_co_propa}a, the lattice period is $i_{1529}=8.3$ \textmu m. At mid-fringe, the clouds are separated by 4.15 \textmu m which is well resolved by our microscope objective.
This allows us to measure the number of transferred atoms $N_{\text{exp}}$ which is obtained by integrating the atomic density over one fringe. 
We additionally compute theoretically the expected atom number $N_{\rm th}$ without any adjustable parameter by integrating the repumped fraction $\rho^{s}_{22}(y,y_r,t_{\rm im})$ {of the strong imaging regime (Eq. (\ref{eq:3level_rhoee}))} over a width of one lattice period:
\begin{equation}
    N_{\text{th}}(y_r) = \frac{P(t_{\text{MW}}) N_{\rm at}^{\rm tot}}{\sqrt{2\pi}\sigma_{y}} \int_{0}^{i_{1529}} \rho^{s}_{22}(y,y_r,t_{\rm im}) dy.
    \label{eq:OneFringeBloch}
\end{equation}

Fig. \ref{fig:atom_number_BAT_co_propa}b shows in situ images of the atom number per pixel for three values of the repumper detuning ($\Delta_{780}=0, U_{\text{5P,0}}/2, U_{\text{5P,0}}$) for an excited state light shift of $U_{\text{5P,0}}=21\Gamma$. {Given that the 1529 nm lattice period is spatially resolved when the detuning is scanned, these images correspond to a tomography of the excited state.} 
Fig. \ref{fig:atom_number_BAT_co_propa}d shows the number of transferred atoms $N_{\text{exp}}$ as a function of the repumper frequency for various excited state light shifts. The points A, B, C correspond respectively to the bottom, middle and top of the 1529 nm lattice as shown on Fig. \ref{fig:atom_number_BAT_co_propa}c. The width from the points A to C is a measure of the light shift and exactly matches with theoretical computations (see Appendix \ref{sec:light_shifts}). 
{Knowing the light shifts and the repumper saturation, the $\rm FWHM_{s}$ (Eq. (\ref{eq:3level_FWHM_limit_middle})) and $N_{\text{th}}$ (Eq. (\ref{eq:OneFringeBloch})) can be straightforwardly computed. An uncertainty of 15$\%$ on $N_{\text{th}}$ originates mainly from the uncertainty on the determination of the total number of atoms $N_{\text{th}}^{\text{tot}}$.}

\begin{figure}
\begin{center}
\includegraphics[scale=0.19]{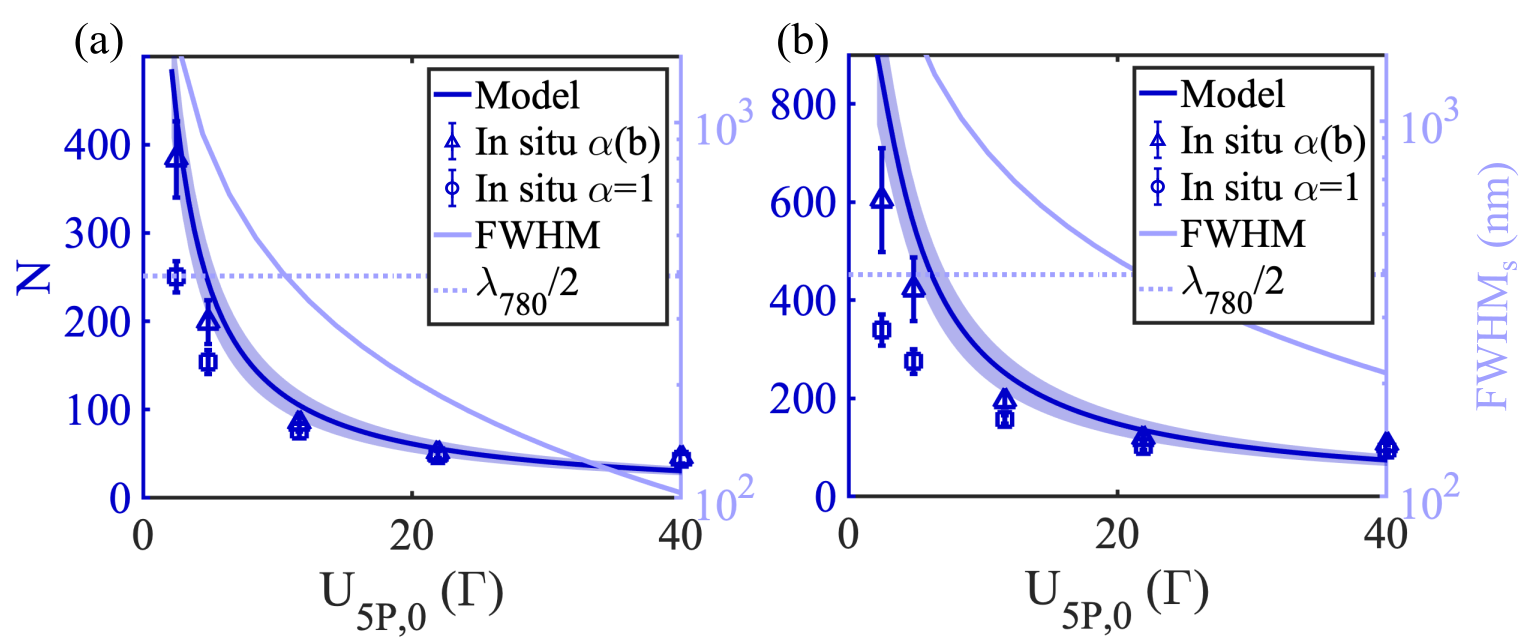}
\caption{Atom number (left axis) per unit of {lattice period} as a function of the light shift on the excited state $\ket{2'}$ when $\Delta_{780}=U_{\text{5P,0}}/2$. The expected atom number is computed from the 3LS model with time-of-flight mesurements (solid line), or in situ with $\alpha=1$ (circles) and $\alpha=\alpha(b)$ (triangles). The FWHM (right axis) corresponds to the full-width at half maximum given by Eq. (\ref{eq:3level_FWHM_limit_middle}). The diffraction limit is shown by the dotted line. The 780 nm laser parameters are (a) $s_0=0.022$, $t=8$ \textmu s and (b) $s_0=0.063$, $t=12$ \textmu s.}
\label{fig:atom_number_middle_co_propa}
\end{center}
\end{figure}

Fig. \ref{fig:atom_number_middle_co_propa} shows the expected atom number $N_{\text{th}}$ and the experimental one $N_{\text{exp}}$ as a function of the light shift amplitude $U_{\text{5P,0}}$ for two repumper saturations. The saturation and duration of repumper have been chosen such that the number of scattered photons on the repumper transition varies from 1 to 10 with either $s_0=0.022$ with $t=8$ \textmu s or $s_0=0.063$ with $t=12$ \textmu s. {For the smallest saturation $s_0=0.022$, the validity criterion in strong imaging $\mathcal{S}$ is higher than 1. It ranges from $460$ ($U_{\text{5P,0}}=2.5\Gamma$) to $1.8$ ($U_{\text{5P,0}}=40\Gamma$). For larger saturations, the adiabacity is more easily fullfilled. In Appendix \ref{subsec:numerical_simulations_Schrodinger}, we checked that the strong imaging regime applies by solving numerically the Schr\"{o}dinger Eq. (\ref{eq:SchrodingerEquation_Main}).} 

In Fig. \ref{fig:atom_number_middle_co_propa}, we see that using the correct atomic scattering cross section ($\sigma_0/\alpha(b)$) leads to a very good agreement between the experimental and theoretical atom numbers $N_{\text{exp}}(\alpha(b))\approx N_{\text{th}}$. In comparison, we show the case of uncorrected data ($\alpha=1$)  and we see that $N_{\text{exp}}(\alpha=1)<N_{\text{th}}$ in the low light shift limit which is where the transferred population $\ket{2}$ is the largest and the multiple scattering effects are the strongest \cite{Veyron2022prr}. For large light shifts ($U_{\rm 5P,0}=40 \Gamma$), we detect more atoms than the model predicts. We attribute this discrepancy to the coupling of the repumper to other hyperfine excited states and to state mixing effects which are not included in the 3LS model. 

The agreement of experimental and theoretical atom numbers confirms the validity of our model. As a result, we show the associated spatial resolutions given by Eq. (\ref{eq:3level_FWHM_limit_middle}) on the right axis of Figs. \ref{fig:atom_number_middle_co_propa}. For a large lattice spacing of $8.3\ \mu m$, we reached a FWHM of the repumped fraction of 100 nm which is smaller than the diffraction limit of $\lambda/2=390$ nm. This resolution being inversely proportional to the lattice spacing, it is expected to gain a factor $8.3/0.77=10.8$ by using counter-propagating beams. 



	
\subsection{{Weak imaging regime with a tightly confined lattice}}\label{sec:1Dlatt_cloud}


We now apply our method to image the longitudinal atomic density of a 1D optical lattice. For that purpose, the atoms are evaporated in a hybrid trap with the single dipole beam DT1 to reach the Bose-Einstein Condensation with $2 \times 10^5$ atoms. After compression of DT1, the trap frequencies are ($15,160,160$) Hz along ($x,y,z$). The atoms are then loaded in a  1064 nm lattice which is adiabatically ramped up from 0 to 40$E_r$, where {$E_r=\hbar^2k_{1064}^2/(2m)$} is the recoil energy at 1064 nm. The lattice depth $U_{\text{5S,0}}$ has been characterized at the atom position using Kapitza-Dirac scattering \cite{Gadway2009,Denschlag2002}. The lattice is compressed up to $1000E_r$ after removing the magnetic gradient and defining the quantization axis via an homogeneous magnetic bias of $280$ mG along $z$. The width of the ground state wavefunction is given by the harmonic oscillator width $a_{\text{HO}}=\sqrt{\hbar/(m\omega_{\text{HO}})}$, where $\omega_{\text{HO}}=2\sqrt{U_{\text{5S,0}}E_r}/\hbar$ is the trap frequency of one site. The standard deviation (std) of the atomic density along $x$ is therefore expected to be equal to $\sigma_x = a_{\text{HO}}/\sqrt{2}=21.2$ nm. Along the $y$-direction, we experimentally measured a width $\sigma_y=$ 6 \textmu m. Because the potential is rotationally symmetric about $x$, we can assume $\sigma_z=\sigma_y$.

\begin{figure}
\begin{center}
\includegraphics[scale=0.37]{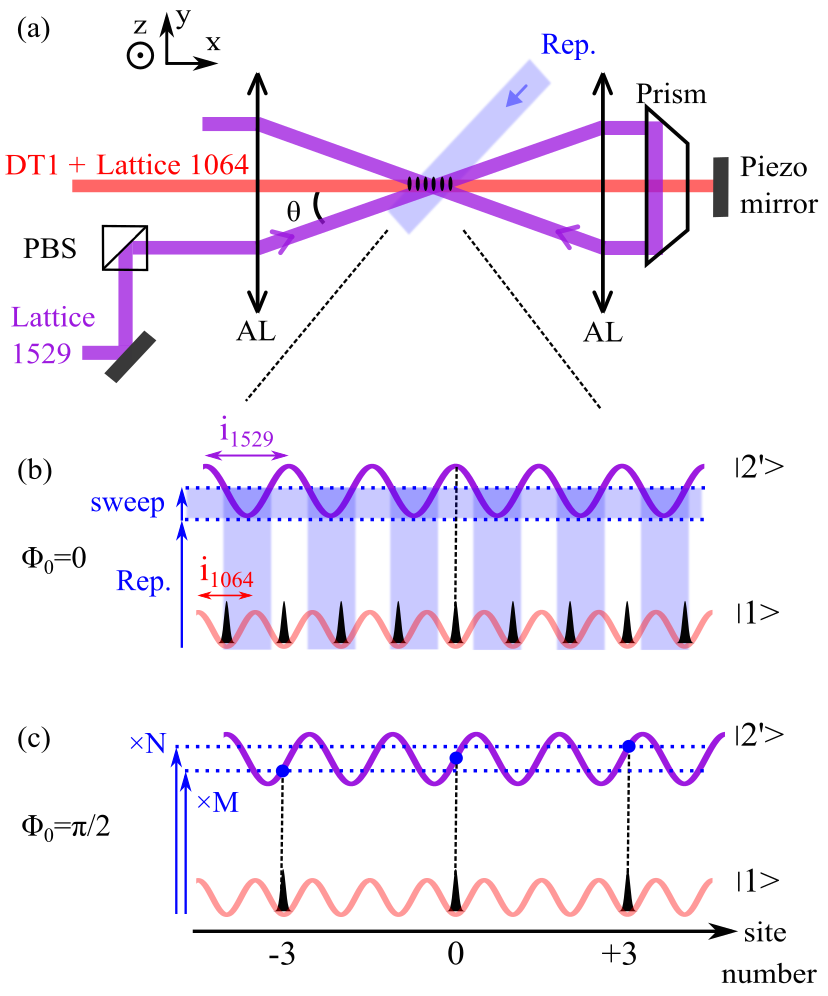}
\caption{(a) Optical setup for the generation of the 1529 nm and 1064 nm lattices. (b) The two lattices with a relative phase $\Phi_0=0$. The blue shaded area corresponds to a  sweep of the repumper frequency during the coarse cleaning of the lattice. (c) The two lattices with a relative phase $\Phi_0=\pi/2$. $M$ (resp. $N$) corresponds to the number of repumper pulses at a specific frequency for the fine cleaning of the -3 (resp. +3) sites.}
\label{fig:counterPropasetup}
\end{center}
\end{figure}

The ground state lattice with a period of $i_{1064}=\lambda_{1064}/2=532.23$ nm is formed by two counter-propagating beams (see Fig. \ref{fig:counterPropasetup}a), yielding a ground state trapping potential given by:
\begin{equation}
\begin{aligned}
U_{\text{5S}}(x) = & U_{\text{5S,0}} \cos^2 \left( \frac{k_{1064}}{2} x + \frac{\Phi_0}{2} \right),
\label{eq:ground_state_lattice}
\end{aligned}
\end{equation}
where $k_{1064}=2\pi/i_{1064}$ is the lattice wavevector and $\Phi_0$ is the relative phase between the 1529 nm and 1064 nm lattices. As shown on Figs. \ref{fig:counterPropasetup}b and \ref{fig:counterPropasetup}c, if $\Phi_0=0$ then both lattice extrema coincide for a central site, while if $\Phi_0=\pi/2$, the same central site is aligned at mid-fringe of the 1529 nm lattice. The relative phase between the two lattices is controlled by a piezo stack that moves the reflecting mirror of the 1064 nm lattice. (see Appendix \ref{sec:calibration_piezo}).

The 1529 nm lattice is generated along the 1064 nm lattice by reflecting a 1529 nm beam with a prism with an angle $\theta$ such that the lattice period is $i_{1529}={\lambda_{1529}}/{(2\cos(\theta))}$. Due to the different 1064 nm and 1529 nm lattice periodicities, a perfect commensurability is obtained for a given number of sites $n_{1064}$ at 1064 nm for the ground state and $n_{1529}$ at 1529 nm for the excited state such that: 
\begin{equation}
    \begin{aligned}
        n_{1529}i_{1529}=n_{1064}i_{1064}.
        \label{eq:CommensurableLattices}
    \end{aligned}
\end{equation}

This gives sets of angles for $\theta$ for which the commensurability is obtained:
\begin{equation}
    \begin{aligned}
        \theta = \acos \left( \frac{n_{1529} \lambda_{1529}}{n_{1064} \lambda_{1064}} \right).
        \label{eq:CommensurableLattices_angle}
    \end{aligned}
\end{equation}

Due to experimental mechanical constraints, the only accessible angle on our setup is $\theta=5.92^{\circ}$ for which we have $n_{1064}=13$ and $n_{1529}=9$. We have therefore a super-lattice period of 6.9 $\mu$m where every 13 periods of the 1064 nm lattice, the atoms will see exactly the same modulation of the 1529 nm lattice in the excited state. This super-lattice period is large and can be resolved easily by our microscope objective which enables the superresolution of the lattice sites. \\

\begin{figure}
\begin{center}
\includegraphics[scale=0.34]{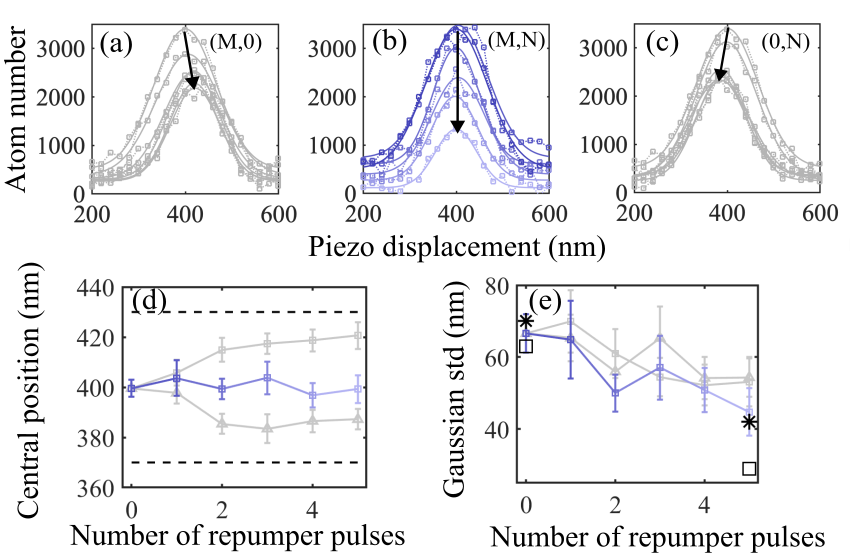}
\caption{(a), (b), (c) Wavepacket density imaging by scanning the piezo mirror. The arrow indicates an increasing number of cleaning pulses of respectively $(M,0)$, $(M,N=M)$ and $(0,N)$ where $M$ and $N$ varies from 0 to 5 pulses. Dotted lines with squares are the data and the lines are Gaussian fits from which the central position (d) and width (e) are extracted. Errorbars are fit errors. In (e), the two squares (resp. stars) show the theoretical widths for $\eta_y=0^{\circ}$ (resp. $\eta_y=0.3^{\circ}$).}
\label{fig:counterPropasetupData}
\end{center}
\end{figure}

To demonstrate the performances of the method, we now aim at measuring the standard deviation of a single site that we will label site number 0. We start by preparing that single site in two steps. In a first step, we perform a coarse cleaning stage in which all  sites except sites 0 and $\pm3$, as indexed in Fig. \ref{fig:counterPropasetup}b, are repumped. For that purpose,  we spatially shift the excited state using a 1529 nm modulation of $U_{\text{5P,0}}=17\Gamma$ and repump the atoms while sweeping in 10 ms the detuning of the repumper from $\Delta_{780}=0.4U_{\text{5P,0}}$ to $U_{\text{5P,0}}$ at a saturation of $s_0=2\times10^{-3}$. All repumped atoms are then pushed away with a resonant $\ket{2}$ to $\ket{3'}$ push laser pulse. In a second step, the piezo is ramped in 100 ms by a distance of $i_{1529}/4$ to align the site 0 onto a slope of the 1529 nm modulation as shown on Fig. \ref{fig:counterPropasetup}c. The sites +3 (resp. -3) are cleaned using $N$ (resp. $M$) short repumper pulses at $\Delta_{780}=3\Gamma$ (resp. $\Delta_{780}=U_{\text{5P,0}}-3\Gamma$). 
Finally, the atomic density is imaged by varying the relative phase $\Phi_0$ and applying a last repumper pulse of detuning $\Delta_{780}=U_{\text{5P,0}}/2$, saturation $s_0=0.02$ and duration $t=16$ \textmu s. {Such parameters correspond to a weak imaging validity criterion. Indeed, using $r_0=0.24$, $\mathcal{W}_2, \mathcal{W}_1= 0.02, 0.03$ which are indeed small compare to 1. In Appendix \ref{subsec:numerical_simulations_Schrodinger}, we verified that the weak imaging regime applies by solving numerically the Schr\"{o}dinger Eq. (\ref{eq:SchrodingerEquation_Main}). Such simulation does not account for the doubly dressed state contribution to the potential \cite{Bellouvet2018} which was shown, in Appendix \ref{sec:forces_excitation}, to have a negligible effect for the considered experimental realisation.} The corresponding atomic densities are shown on Fig. \ref{fig:counterPropasetupData}a,b,c. Each curve is fitted by a Gaussian function from which the central position and std are shown on Fig. \ref{fig:counterPropasetupData}d,e. \\


As $M$ increases while $N=0$ (Fig. \ref{fig:counterPropasetupData}a), cleaning only the sites $-3$ shifts the wavepacket to the right after which it remains stable. In that case we can consider that the $-3$ site is empty. As the site $0$ is centered at the middle of the excited state modulation, the sites $\pm3$ would be resonant exactly at $(3i_{1064}-2i_{1529})=60$ nm away from that central position. Therefore, for equal initial populations in the sites $-3,+3,0$, the central position should shift by at maximum $30$ nm. However, during the coarse cleaning stage, the $\pm3$ sites are closer from resonance than the site $0$ and the repumper therefore induces more scattering on these sites and reduces more their population. These unequal populations would lead to a smaller shift of the central position. From numerical simulations (see Appendix \ref{sec:central_position_shifts}), the measured experimental shift of about 20 nm is reproduced for relative populations of 0.6 in the sites $\pm3$ compared to the population in the site 0. 

As both $M$ and $N$ increase, the Gaussian std decreases as expected as $\pm3$ sites get cleaned (Fig. \ref{fig:counterPropasetupData}e). At minimum, the expected std is {29 nm which corresponds to the width numerically obtained from Eq. (\ref{eq:FWHMweak_main}).} We measured a std of $45\pm5$ nm which is slightly larger than the expected one. This expected width limit (square points) corresponds to the case of perfect alignment of the two lattices. It is however very sensitive to the relative angle between the lattice wavevectors. In Appendix \ref{sec:angle_sensitivity}, we computed the effective std $\tilde{\sigma}_x$ after adding a rotation by an angle $\eta_y$ about the $y$-axis between the two lattices. We showed that in the small angle limit the std is given by $\tilde{\sigma}_x=\sigma_x\sqrt{1+\eta_y^2(\sigma_y/\sigma_x)^2}$. For only $0.3^{\circ}$ (star points in Fig. \ref{fig:counterPropasetupData}e), the calculated widths overlap with the experimental data.


\section{Conclusion}
{In this work, we have super-resolved atomic position using the high sensitivity of atomic properties to a spatially varying electromagnetic environment. In particular, we have demonstrated that excited state energy shifts were a versatile and well controlled solution to manipulate atomic transition frequencies over very short distances. We have given a general theoretical framework to describe subwavelength imaging dynamics and derived the expected localization performance for two opposite regimes of imaging strength for which we have determined validity criteria. Both regimes lead to conceptually very different situations but achieve similar resolution capabilities.
Experimentally, in \ref{sec:ho_cloud} we first applied our excited state engineering method in the strong imaging regime. }Imaging optically resolved slices of atoms, we have demonstrated an excitation length around 100 nm. This resolution is well explained and quantitatively corresponds to a model accounting for the internal state dynamics of a 3LS. We emphasize that the excitation length estimate relies on an absolute measurement of atom numbers. The quantitative match between the model and experimental data for large optical depths (low modulation depths) requires to account for a non-negligible reduction of the scattering cross section that was characterized in \cite{Veyron2022prr}.{ For small excitation lengths, the effect of kinetic energy could be mitigated by deeply entering into the strong imaging regime (shorter time and higher tranfer rate). In the non-optically resolved limit (\ref{sec:1Dlatt_cloud}), we have shown that the resolution could be strongly improved using excited state energy shifts varying at the scale of the wavelength. These shifts were created using counter-propagating laser fields. Such experiment was performed in the opposite weak imaging regime. The method was used to image a strongly compressed atomic density of std $21$ nm and obtained an image of std $45\pm5$ nm, well below the diffraction limit. The increase is attributed both to the finite resolution and to a possible residual fringe misalignment.} 



{One should note that the proposed subwavelength imaging method mostly  addresses the part of the cloud which is finally imaged, leaving the rest of the cloud  little perturbed. As such, it can be used to shape and quench the dynamics of the system. Regarding the practical implementation, non periodic transfer could be reached using structured light excited state dressing that would further enable a parallelized and tunable imaging of multiple slices. As compared to \cite{McDonald2019,Subhankar2019}, the position selectivity of the presented method also depend on the standing waves positions but can additionally be fine tuned by the adjustment of the repumping radiation frequency. 
The  repumping radiation power directly affects the on-resonant scattering rate that sets the imaging strength and allows to scan from the weak to the strong imaging regime.}

The current experiments and numerical simulations have been performed for $^{87}$Rb atoms. However, it can be straightforwardly extended to other alkali metals with large excited state hyperfine splittings such as Cesium atoms. For alkalis with lower excited state splittings, the method could straightforwardly be extended to the large field limit which only results in a redefinition of the proper eigenbasis. 
{In our imaging system, we used fast absorption imaging to infer atom numbers. Single-atoms fluorescence detection \cite{Fuhrmanek2011,Schaefer2020} could be straightforwardly implemented to reach single-atom sensitivity by reading out the hyperfine state $\ket{2}$.}
{With our method, the resolution depends both on the excited state shift and on the excited state linewidth, making it favorable for atomic species with narrow transitions. For instance, spectrally-resolved imaging has been demonstrated using an ultranarrow optical transition in Ytterbium atoms \cite{Shibata2014}.} Other examples of usable narrow optical transitions include Strontium and Dysprosium species. {For such narrow transitions ($\Gamma$ small), the strong imaging regime criterion ($\mathcal{S}\gg1$) can hardly be reached as $\mathcal{S}\propto \Gamma^3$ while the weak imaging regime criterion ($\mathcal{W}\ll1$) is favored as $\mathcal{W}\propto \Gamma^2$ but hardly allows to image the state with high temporal resolution.}
{We finally emphasize that our method could be spin-selective by taking advantage of the differential light shifts between Zeeman states for circularly polarized 1529 nm beams.}

\begin{acknowledgments}
 R.V. acknowledges PhD support from the University of Bordeaux,  J-B.G. and V.M. acknowledge support from the French State, managed by the French National Research Agency (ANR) in the frame of the Investments for the Future Programme IdEx Bordeaux-LAPHIA (ANR-10-IDEX-03-02). This work was also supported by the ANR contracts (JCJC ANR-18-CE47-0001-01 and QUANTERA21 ANR-22-QUA2-0003) and the Quantum Matter Bordeaux.
\end{acknowledgments}

\FloatBarrier
\section*{Appendix} \label{sec:appendix}

{In the Appendix \ref{sec:light_shifts}, we compare the experimental light shifts induced by the 1529 nm laser to theoretical computations. In Appendix \ref{sec:Rb87_D2_3LS}, we justify the reduction to a three-level system that is used to model the internal state transfer dynamics. In Appendix \ref{sec:ST_dynamics}, we describe the spatio-temporal dynamics during the transfer, the weak and strong imaging regimes with analytical criteria for their domain of validity, and simulate the full system for the experiments presented in the manuscript. In Appendix \ref{sec:forces_excitation}, we discuss the impact of the light shift gradient of the excited state for the experimental conditions. In Appendix \ref{sec:calibration_piezo}, we show how the piezo stack controlling the 1064 nm is calibrated. Finally, we simulate the wavepacket microscopy sequence in the 1064 nm lattice to study the influence of experimental imperfections on the central position of the wavepacket in Appendix \ref{sec:central_position_shifts} and its associated width in Appendix \ref{sec:angle_sensitivity}.}

\appendix
\section{Light shifts at 1529 nm}\label{sec:light_shifts}
The light shifts induced by the 1529 nm laser are computed by diagonalising the total Hamiltonian composed of the AC Stark Hamiltonian and the hyperfine Hamiltonian \cite{Steck2019book}. 
To compute the light shifts for a far-off-resonance laser, the counter-rotating term of the atom-field interaction has to be included. As both terms oscillate rapidly compared to each other, both light shifts computed independently from each other can be added after their respective diagonalisation in their own rotating-frame.

The numerical computations of the light shifts of the state $5P_{3/2}$ with a laser at exactly 1529.36098 nm includes the following states: $5S_{1/2}$, $5P_{1/2}$, $5P_{3/2}$, $6P_{1/2}$, $4D_{3/2}$ and $4D_{3/2}$. The main transitions that contribute to the light shifts are $5P_{3/2}$ to $4D_{5/2}$ at 1529.366 nm with a dipole moment of $10.899ea_0$, and $5P_{3/2}$ to $4D_{3/2}$ at 1529.262 nm with a dipole moment of $3.628ea_0$, where $e$ the electron charge and $a_0$ the Bohr radius. Typical atomic parameters for $^{87}$Rb can be found in \cite{Arora2019,Arora2007}. Our 1529 nm laser mainly drives the transition between $5P_{3/2}$ and $4D_{5/2}$ as depicted on Fig. \ref{fig:3levelAtomDiagram_5levelAtomDiagram}a. The theoretical light shift is ultimately obtained by the knowledge of the intensity of the 1529 nm laser beams {which are uniform over the cloud widths.} {Individual waists of 131 and 138 \textmu m are measured using a tomography technique \cite{Bertoldi2010}. The intensity is controlled by the locked optical power and yields for instance a peak intensity of $4.1\times10^5$ W/m$^2$ for an optical power per beam of 2.9 mW.} The 1529 nm resonance is found by a spectroscopic scan onto the atoms.

\begin{figure}
\begin{center}
\includegraphics[scale=0.35]{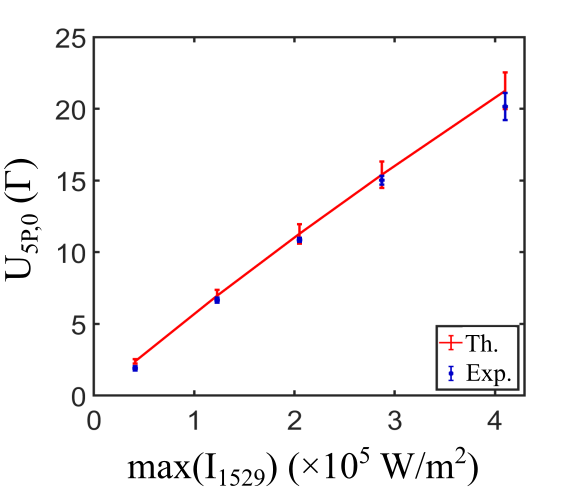}
\caption{Experimental light shifts obtained from tomographic measurements compared to the theoretical shifts computed from the knowledge of the waists, frequency and powers of the 1529 nm beams. Errorbars correspond to the waist uncertainty for the theoretical model and to the standard deviation over 5 realizations for the experimental data.}
\label{fig:lightshift_comparison}
\end{center}
\end{figure}

We compared this theoretical description of the light shift with the experimental realizations on Fig. \ref{fig:lightshift_comparison} for the homogenous cloud configuration. 5 repetitions of a tomographic curve are measured, from which the light shifts are obtained by fitting the atom number with Eq. (\ref{eq:3level_rhoee_remind}). Both agree which enables us to precisely compute the spatial resolution with Eq. (\ref{eq:3level_FWHM_limit_middle}) where the light shift is an input parameter.

\section{3LS repump model for $^{87}$Rb}\label{sec:Rb87_D2_3LS}
\begin{figure}
\begin{center}
\includegraphics[scale=0.3]{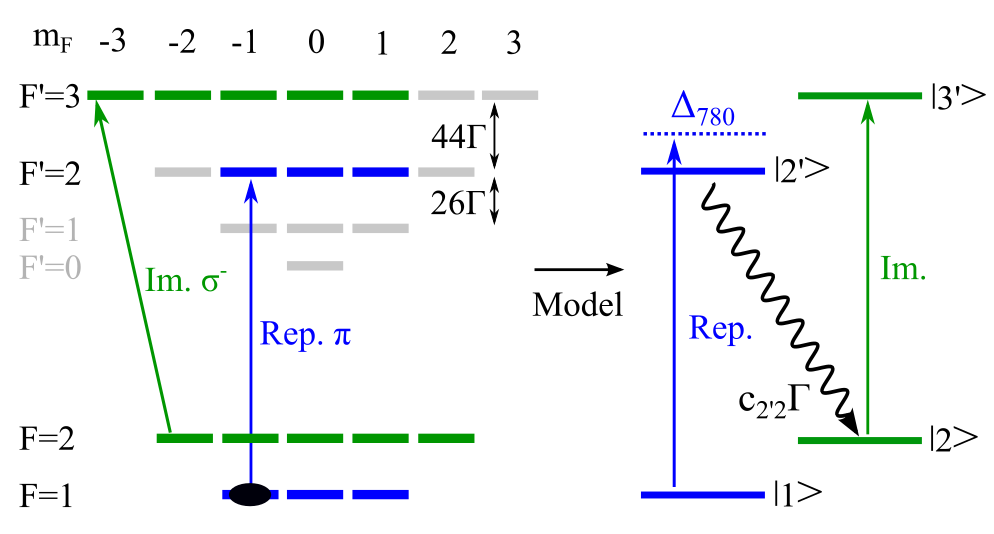}
\caption{$^{87}$Rb D$_2$ transition hyperfine structure ($5^2S_{1/2}$ to $5^2P_{3/2}$) and its approximation as a 3LS for the repumper transition and as a 2LS for the imaging transition.}
\label{fig:appendix_all_levels}
\end{center}
\end{figure}

Fig. \ref{fig:appendix_all_levels} shows the effective 3LS that we consider to model the point spread function based on the multi-level structure of $^{87}$Rb. 

The differential light shifts between the $m_F$ states of $F'=2$ for a $\pi$-polarized 1529 nm laser are much lower than the atomic linewidth $\Gamma/2\pi=6.066$ MHz. As a result, there is no dark state in the hyperfine ground state $F=1$ as all $m_F$ states can be locally coupled to the excited state. We numerically checked that taking all the states coupled by $\pi$-transitions or approximating it as a two-level system leads to the same population transfer. Therefore, we treat the multi-level repumper transition as the two-level system: $\ket{1}$ and $\ket{2'}$ (blue levels in Fig. \ref{fig:appendix_all_levels}). For this two-level system, as the atom is initially prepared in $\ket{5^2S_{1/2},F=1,m_F=-1}$, we include in the saturation parameter the coupling strength of the transition between $\ket{5^2S_{1/2},F=1, m_F=-1}$ and $\ket{5^2P_{3/2},F'=2, m_F=-1}$ of $-\sqrt{1/8}$ \cite{SteckRb872001} which leads to a saturation intensity of $I_{\text{sat,rep}}=6.67$ mW/cm$^{2}$. 

We ignore the coupling from $\ket{1}$ to the excited state $\ket{5^2P_{3/2},F'=1,m_F=-1}$ as it is $26\Gamma$ away from $\ket{2'}$ which is smaller that the 1529 nm induced light shifts of the experiment. 

The population in $\ket{1}$ is transferred via absorption/spontaneous emission cycles into $F=2$ and remains there: we treat all $m_F$ states of $F=2$ as a single state $\ket{2}$. 

Finally, the population in $\ket{2}$ is measured by absorption imaging using a circularly polarized laser tuned on the cycling transition from $\ket{5^2S_{1/2},F=2,m_F=-2}$ to $\ket{5^2P_{3/2},F'=3,m_F=-3}$.

\section{Spacio-temporal dynamics during imaging.}\label{sec:ST_dynamics}

\subsection{Temporal dynamics}\label{subsec:optical_Bloch_eq_dynamics}
{The point spread function (PSF), defined in our system as the spatially dependent population transfer rate to $\ket{2}$ ($\kappa(x)$) is derived from the time evolution of the density matrix $\rho$ of the 3LS in abscence of spatial dynamics \cite{Steck2019book}.}

{We focus on slow dynamics ($\Gamma t\gg1$) by applying the adiabatic approximation which assumes that the optical coherences are always in equilibrium with respect to the populations ($\dot{\rho}_{12'}\approx0$). This regime is valid for thermal atoms if their displacement over a time $1/\Gamma$ is much smaller than the target subwavelength resolution. Atoms at 169 nK (section \ref{sec:ho_cloud}) move by 170 pm during the 28 ns of the scattering event which is indeed much smaller than the target resolution of 10 nm. In this case, the 3LS density matrix evolution simplifies to rate-equations:
\begin{equation}
\frac{d}{dt} \begin{pmatrix} 
{\rho_{11}}  \\ {\rho_{2'2'}} \\  {\rho_{22}}  \\    
\end{pmatrix}
= 
\begin{pmatrix}  
-\frac{s\Gamma}{2} & \frac{s\Gamma}{2} + c_{2'1}\Gamma & 0 \\ 
\frac{s\Gamma}{2} & -\frac{s\Gamma}{2} - \Gamma & 0 \\ 
0 & c_{2'2}\Gamma & 0 \\ 
\end{pmatrix}
\begin{pmatrix} 
\rho_{11} \\ \rho_{2'2'} \\  \rho_{22} \\    
\end{pmatrix},
\label{eq:3level_matrix_RateEquations}
\end{equation}
where $c_{2'1}=1/2$ (resp. $c_{2'2}=1/2$) are the normalized decay rates from $\ket{2'}$ to $\ket{1}$ (resp. $\ket{2}$) due to spontaneous emission at a rate $\Gamma$, and $s(x)=s_0/(1+(2\Delta(x)/\Gamma)^2)$.}


{An analytical solution of Eq. (\ref{eq:3level_matrix_RateEquations}) is obtained by diagonalisation method. Starting from the initial state $\rho_{11}(t=0)=1$, for any saturation parameter the populations are equal to:
\begin{equation}
    \begin{aligned}
        &\rho_{11}  = 1-\rho_{2'2'}-\rho_{22}, \\
        &\rho_{2'2'}  = \frac{\Gamma+\Lambda_{-}+\Lambda_{+}}{2(\Lambda_{-}-\Lambda_{+})} \left( e^{\Lambda_{+}t}-e^{\Lambda_{-}t} \right), \\
        &\rho_{22}  = 1-\frac{\Lambda_{-}}{\Lambda_{-}-\Lambda_{+}}e^{\Lambda_{+}t}+\frac{\Lambda_{+}}{\Lambda_{-}-\Lambda_{+}}e^{\Lambda_{-}t},
        \label{eq:SolutionRateEquations_3levels}
    \end{aligned}
\end{equation}   
where $\Lambda_{\pm}=-\frac{\Gamma}{2}\left(1+s \pm \sqrt{1+2sc_{2'1}+s^{2}}\right)$ correspond to non-zero eigenvalues.}

{The $e^{\Lambda_{+}t}$ terms decay to zero at a rate of $\Gamma$ so the long time dynamics is mainly given by the $e^{\Lambda_{-}t}$ terms. The solutions in the low saturation limit $s_0\ll1$ and for $\Gamma t\gg1$ simplify to:
\begin{equation}
    \begin{aligned}
        \rho_{11}(x) & \approx 1-\rho_{22}(x),\\
        \rho_{2'2'}(x) & \approx \frac{1}{2} \frac{s_0}{1+(2\Delta(x)/\Gamma)^2} e^{- \frac{c_{2'2}\Gamma t}{2} \frac{s_0}{1+(2\Delta(x)/\Gamma)^2}},\\
        \rho_{22}(x) & \approx 1 - e^{- \frac{c_{2'2}\Gamma t}{2} \frac{s_0}{1+(2\Delta(x)/\Gamma)^2}}.
        \label{eq:3level_rhoee_remind}
    \end{aligned}
\end{equation}}


{The population in $\ket{2}$ can be reinterpreted as a depumping from $\ket{1}$ with a rate:
\begin{equation}
\begin{aligned}
\kappa({x}) = \frac{c_{2'2}s_{0}\Gamma}{2} \frac{1}{1+(2\Delta(x)/\Gamma)^2}= \frac{\kappa_0}{1+\left(\frac{x}{X_0}\right)^{2}} ,
\label{eq:dissipation_equation}
\end{aligned}
\end{equation}
where $\kappa_0=c_{2'2} s_0 \Gamma/2$ is the on-resonnance transfer rate and $X_0=i_{1529}\Gamma/(2\pi U_{5P,0})$ is the characteristic sub-wavelength length scale that was obtained by linearising $\Delta({x}) = \frac{U_{5P,0}}{2\Gamma} \sin(k_{1529}{x})$ around $x=0$ (mid-fringe).}

{It is interesting to mention that FWHMs can be derived for any resonance condition starting from the expression of $\rho_{22}(x)$. For instance, when the repumper laser is tuned at the bottom of the modulation where $\Delta_{780}=0$, the FWHM in the strong imaging regime is:
\begin{equation}
\begin{aligned}
\text{FWHM}^{\rm bottom}_{\text{s}} = \left( \frac{2\Gamma}{U_{\text{5P,0}}} \right)^{\frac{1}{2}} \frac{i_{1529}}{\pi} \left(  \frac{c_{2'2} \Gamma s_{0} t}{2  \ln(2)}  \right)^{\frac{1}{4}}.
\label{eq:3level_FWHM_limit_bottom}
\end{aligned}
\end{equation}
This configuration is relevant for probing the site occupancy rather that the intra-site details.}

\subsection{Spatio-temporal dynamics}
{The strong localization of particles induced by the imaging process translates in rapid spatial variation of the density profile. It corresponds to a high kinetic energy term that influences the imaging dynamics itself. To phenomenologically accounts for this effect, we model the depumping from state $\ket{1}$ by a non-Hermitian loss term in the Schr\"{o}dinger equation :
\begin{equation}
    \begin{aligned}
    i\hbar\frac{d\ket{\Psi(x,t)}}{dt} = \hat{H}_{\text{0}}\ket{\Psi(x,t)} - i\frac{\hbar\kappa(\hat{x})}{2}\ket{\Psi(x,t)}
    \label{eq:SchrodingerEquation}
    \end{aligned}
\end{equation}
where $\hat{H}_{\text{0}}$ is the Hamiltonian including kinetic and potential energies of a harmonic oscillator with eigenstates $\ket{n}$ and eigenvalues $E_n$ such that $\hat{H}_{\text{0}}\ket{n}=E_n\ket{n}$. $\ket{n}$ corresponds to the harmonic oscillator wavefunction given by $\Phi_n(x)=\braket{x}{n}$.
In Eq. (\ref{eq:SchrodingerEquation}), we have neglected the interactions between particles. Indeed, in the first experiment, the atoms are in a shallow trap which ensures weak interactions. And in the second experiment, a 1D regime applies along the deep lattice direction.
Eq. (\ref{eq:SchrodingerEquation}) is valid for both the strong and weak imaging regimes. A numerical integration is possible for any set of experimental parameters.}

{In the next appendices, we present the simulations carried out for the exprimental situations encountered in the main text. To gain in generality, we algo algebraically derive the criteria that correspond to the weak and strong imaging limit.}

\subsection{Strong imaging criteria}\label{an:strongimaging}
{In the strong imaging regime, we imprint a dip in $\Psi(x,t)$. This generates a velocity spread $\Delta v$. In that case, one should satisfy $\Delta v.t_{\rm imaging}\ll \Delta x$, where $\Delta x$ is the position spread (STD).}

{In the main text, we have discussed the simple and analytical case of a Gaussian dip that satisfies the Heisenberg equality. The opposite limit is represented by the steep gate function of width $L$ for which the velocity spread is infinite. To better model the real probability distribution (Eq. (\ref{eq:3level_rhoee_remind})) whose velocity spread is not analytic, we approximate $\rho_{22}$ by a $C^1$ analytic solution ($\rho_{22}^{\rm approx}$) with conserved amplitude and slope at half-width at half maximum compared to Eq. (\ref{eq:3level_rhoee_remind}):
\begin{eqnarray}
\begin{aligned}
\rho_{22}^{\rm approx}=\frac{A}{2} \left\{ 
    \begin{array}{ll}
         1+\sin(\alpha(x+x_0)) & \mbox{, } |x+x_0| <\frac{\pi}{2\alpha}\\
        2 & \mbox{, } |x|<x_0-\frac{\pi}{2\alpha} \\
       1+\sin(\alpha(x-x_0))
        & \mbox{, } |x-x_0|<\frac{\pi}{2\alpha} \\
        0 & \mbox{otherwise }
    \end{array}
    \right.
\end{aligned}
\label{eq:approx_wavefunction}
\end{eqnarray}
with $A=\frac{1-\exp{-\kappa_0 t}}{\rm FWHM} \approx \frac{1}{\rm FWHM}$ in the deep depletion limit ensures that $\int{\rho_{22}(x)dx}=1$, $x_0=\frac{\rm FWHM}{2}$ and $\alpha=\frac{4 \ln{2}^{3/2}}{X_0 \sqrt{\kappa_0 t}}$. \\
The kinetic energy of a particle of mass $m$ in such wavefunction $\Psi_{22}=\sqrt{\rho_{22}^{\rm approx}(x)}$ can be defined as: 
\begin{eqnarray}
\begin{aligned}
    E_c&=\frac{m \Delta v ^2}{2} =\frac{\hbar^2}{2 m}\int{\Psi_{22}^{\rm approx}(x)\Delta \Psi_{22}^{\rm approx}(x) dx}\\
    &=\frac{\alpha}{8 \rm FWHM}
    \label{eq:Kinetic}
\end{aligned}
\end{eqnarray}
}

{Therefore, a Lorentzian pumping scheme leads to a velocity spread expression:
\begin{eqnarray}
\begin{aligned}
\Delta v&=&\frac{\hbar \sqrt{\ln{2}}}{m \rm FWHM}
\label{eq:vel_spread}
\end{aligned}
\end{eqnarray}
Eq. (\ref{eq:vel_spread}) is used to compute the condition on the pumping strength given by Eq. (\ref{eq:Heisenberg2}) in the main text.}

\subsection{Weak imaging criteria}\label{subsec:calibration_piezo}
{The Schr\"{o}dinger equation (\ref{eq:SchrodingerEquation}) has an adiabatic evolution with respect to the motional states of the harmonic oscillator if the coupling strength satisfies:
\begin{equation}
    \begin{aligned}
    |\bra{\phi_n}\kappa(\hat{x})\ket{\phi_0}| \ll \frac{2(E_n-E_0)}{\hbar}.
    \label{eq:adiabaticity_criteria}
    \end{aligned}
\end{equation}
where the factor of 2 on the right hand side comes from the dissipation term $\kappa(\hat{x})/2$ in the Schr\"{o}dinger equation.\\
It can be explicitly evaluated using the harmonic oscillator eigenfunctions given by the Hermite polynomials $H_n$:
\begin{equation}
    \begin{aligned}
    \braket{x}{\phi_n} = \sqrt{\frac{1}{2^n n!}} \left(\frac{1}{\pi {a_{\text{HO}}}^2}\right)^{1/4} e^{-\frac{x^2}{2{a_{\text{HO}}^2}}} H_n\left(\frac{x}{a_{\text{HO}}}\right)
    \label{eq:QHO_functions}
    \end{aligned}
\end{equation}
where $a_{\text{HO}}$ is the harmonic oscillator width.\\
Using the expression of the dissipation $\kappa({\hat{x}})$, we get:
\begin{equation}
    \begin{aligned}
    \bra{\phi_n}\kappa(\hat{x})\ket{\phi_0} =  \frac{\kappa_0}{\sqrt{\pi2^n n!}} \int_{-\infty}^{+\infty} \frac{dx}{a_0} \frac{H_n\left(\frac{x}{a_0}\right)e^{-\frac{x^2}{a_0^2}}H_0\left(\frac{x}{a_0}\right)}{1+\left(\frac{x}{X_0}\right)^2 }
    \label{eq:adiabaticity_criteria2}
    \end{aligned}
\end{equation}
After a change of variable $x=a_{\text{HO}}X$, and using $H_0(X)=1$, it simplifies as:
\begin{equation}
    \begin{aligned}
    \bra{\phi_n}\kappa(\hat{x})\ket{\phi_0} =  \frac{\kappa_0}{\sqrt{\pi 2^n n!}} \int_{-\infty}^{+\infty} dX \frac{H_n(X)e^{-X^2}}{1+\left(\frac{X}{r_0}\right)^2 }
    \label{eq:adiabaticity_criteria2b}
    \end{aligned}
\end{equation}
where $r_0=X_0/a_{\text{HO}}$.\\
So far, this criterion corresponds to the case where the dissipation operator is centered with respect to the harmonic oscillator. Including a position offset $x_r$ for the dissipation, the matrix element is written more generally as:
\begin{equation}
\begin{aligned}
\bra{\phi_n}\kappa(\hat{x}-x_r)\ket{\phi_0} =  \frac{\kappa_0}{\sqrt{\pi 2^n n!}} \int_{-\infty}^{+\infty} dX \frac{H_n(X)e^{-X^2}}{1+\left(\frac{X}{r_0}-\frac{x_r}{X_0}\right)^2 }
\label{eq:adiabaticity_criteria3}
\end{aligned}
\end{equation}
Eq. (\ref{eq:adiabaticity_criteria3}) gives the matrix element without approximation for any position $x_r$. In the following, we evaluate it in the strong localization limit ($r_0\ll 1$) to derive simple analytical criteria for the adiabaticity, and give a resolution for the case $x_r=0$ beyond this limit.}

\subsubsection{Strong localization limit}
{From Eq. (\ref{eq:adiabaticity_criteria3}), in the limit of strong localization where $r_0\ll 1$, we can treat the Lorentzian function as a Dirac distribution: 
\begin{equation}
     \lim_{r_0\to 0}  \frac{1}{1+\left[\frac{X}{r_0}-\frac{x_r}{X_0}\right]^2}  = \pi r_0 \delta\left(X-\frac{x_r}{a_\text{HO}}\right),
\end{equation}
which yields
\begin{equation}
    \bra{\phi_n}\kappa(\hat{x}-x_r)\ket{\phi_0} = \kappa_0 r_0 \sqrt{\frac{ \pi}{2^n n!}}  H_n\left(\frac{x_r}{a_\text{HO}}\right)e^{-\left(\frac{x_r}{a_\text{HO}}\right)^2}.
\end{equation}
Maximizing this matrix element with respect to the repumping position $x_r$ is equivalent to finding the smallest root $x_n$ of $H_{n+1}$, as it can be shown using Hermite polynomials recursion relations for $H'_n(x)$. For $n \geq 2$ even, the matrix element is maximum at $x_n=0$, while for $n \geq 1$ odd, it is maximum at the first antinode of $H_{n}$. There is unfortunately no formula for the position of those roots for any $n$ so we found them algebraically for $n=1,2$ for which $x_1 = 1/\sqrt{2}$ and $x_2 = 0$. We also verified numerically that $ \mathcal{W}_n = \bra{\phi_n}\kappa(\hat{x}-x_r)\ket{\phi_0}/n$ is monotonically decreasing with $n$ and that the adiabatic condition is fulfilled in the strong localization limit:
\begin{equation}
    \begin{aligned}
        &\mathcal{W}_1 =\sqrt{\pi}e^{-\frac{1}{2}} \frac{\kappa_0 r_0}{2\omega_{\text{HO}}} + o(r_0)  \ll 1, \\
    &\mathcal{W}_2 =\sqrt{\frac{\pi}{8}} \frac{\kappa_0 r_0}{2\omega_{\text{HO}}} < \mathcal{W}_1, \\
    &\mathcal{W}_{n+1}<\mathcal{W}_{n} \quad \forall n.
    \label{eq:adiabaticity_criteria_n_is_1_largelimit}
    \end{aligned}
\end{equation}
}

\subsubsection{Beyond the strong localization limit}
{Beyond the strong localization limit, it is possible to get analytical results for the matrix element in the case $x_r=0$.\\
For $n$ odd, the Hermite polynomials are odd so $\bra{\phi_n}\kappa(\hat{x})\ket{\phi_0}=0$. It is due to the fact that the dissipation is centered about the harmonic oscillator center. Therefore, an odd state cannot be coupled from an even state. Odd states could be coupled in the case where there would be a spatial offset between the dissipation and the harmonic oscillator.\\
For $n$ even, $\bra{\phi_n}\kappa(\hat{x})\ket{\phi_0}$ is non-zero and can be computed using the general polynomial expansion of the Hermite functions:
\begin{equation}
    \begin{aligned}
        H_n(X) = n! \sum_{m=0}^{\lfloor\frac{n}{2}\rfloor} \frac{(-1)^m}{m!(n-2m)!}(2X)^{n-2m},
        \label{eq:series_Hermite_poly}
    \end{aligned}
\end{equation}
where $\lfloor.\rfloor$ denotes the floor function.\\
Using Eq. (\ref{eq:series_Hermite_poly}), the coupling term $\bra{\phi_n}\kappa(\hat{x})\ket{\phi_0}$ for even $n$ reads:
\begin{equation}
    \begin{aligned}
        \bra{\phi_n}\kappa(\hat{x})\ket{\phi_0} = &\frac{\kappa_0 r_0^2e^{-r_0^2}}{\sqrt{\pi2^n}} \sqrt{n!} \sum_{m=0}^{n/2} (-1)^m \frac{ 2^{n-2m}}{m!(n-2m)!} \times \\ 
        &  \Gamma_f\left(\frac{1}{2}(1-2m+n)\right) \text{Ei}_{\frac{1}{2}(1-2m+n)}\left(r_0^2\right)
        \label{eq:coupling_neven_0}
    \end{aligned}
\end{equation}
where $\Gamma_f$ is the Gamma function and $\text{Ei}$ is the generalized exponential integral function.\\
From this result, we can extract an exact analytical expression for the adiabaticity criterion for any even quantum number $n$. The most stringent adiabatic criterion given by Eq. (\ref{eq:adiabaticity_criteria}) is obtained for $n=2$:
\begin{equation}
    \begin{aligned}
        \abs{ \frac{\kappa_0 r_0^2e^{r_0^2}}{2\sqrt{2}} \left( \text{Ei}_{\frac{1}{2}}\left(r_0^2\right) - \text{Ei}_{\frac{3}{2}}\left(r_0^2\right) \right)  } \ll 2\omega_{\text{HO}}
        \label{eq:adiabaticity_criteria_stringent}
    \end{aligned}
\end{equation}
Equivalently, Eq. (\ref{eq:adiabaticity_criteria_stringent}) can be written as:
\begin{equation}
    \begin{aligned}
        \abs{ \frac{\kappa_0 r_0}{2\sqrt{2}} \left( 2r_0 - (1+2r_0^2)e^{r_0^2}\sqrt{\pi} \text{erfc}(r_0) \right)  } \ll 2\omega_{\text{HO}}
        \label{eq:adiabaticity_criteria_n_is_2}
    \end{aligned}
\end{equation}
where $\text{erfc}$ is the complementary error function.\\
In the regime of strong localisation of the wavefunction where $r_0\ll 1$, the criterion leads to $\mathcal{W}_2$ of Eq. (\ref{eq:adiabaticity_criteria_n_is_1_largelimit}).}





\subsection{Numerical simulation of the evolution} \label{subsec:numerical_simulations_Schrodinger}
\begin{figure}
\begin{center}
\includegraphics[scale=0.17]{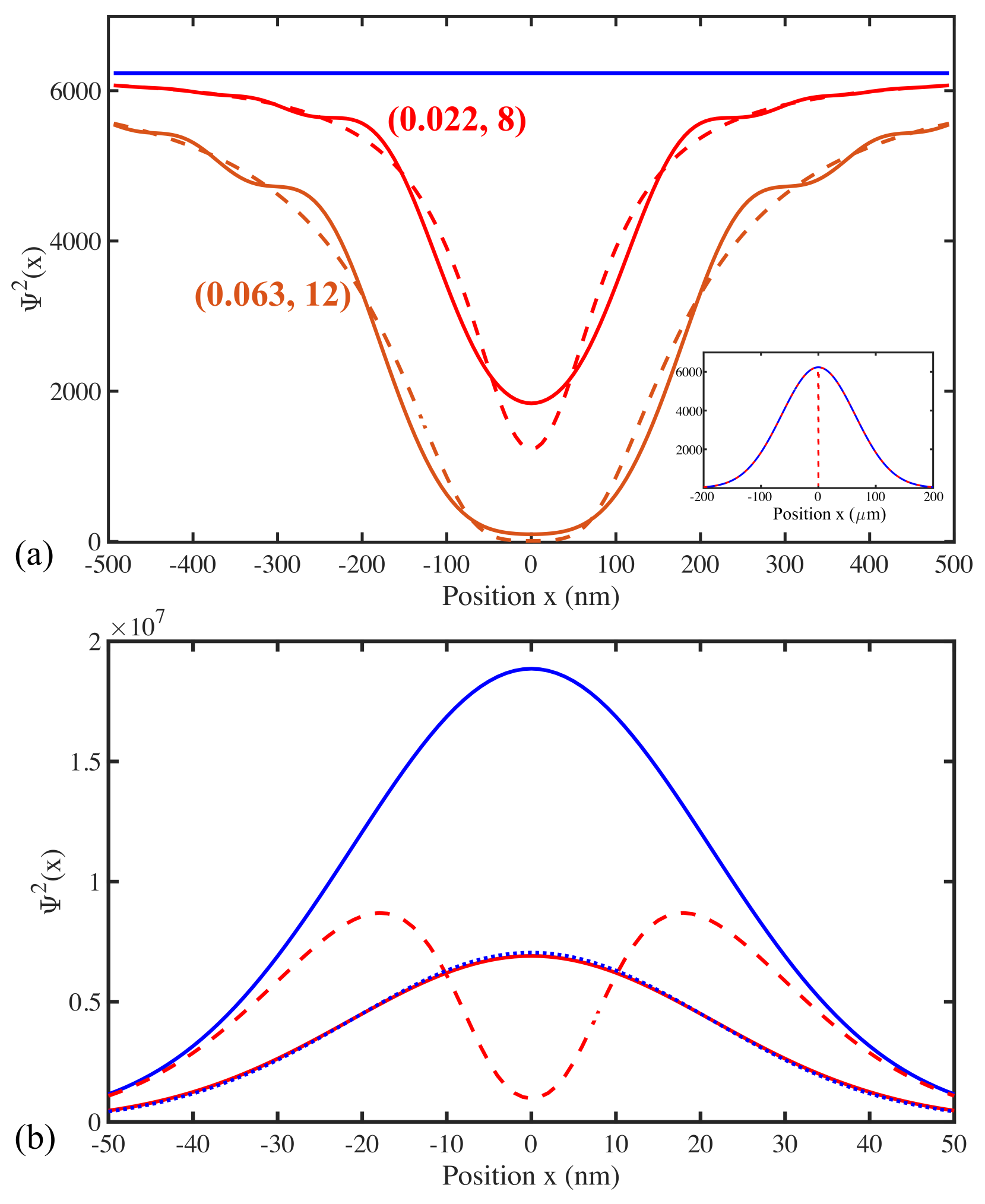}
\caption{Simulation of the evolution of the wavefunction. The solid blue lines correspond to the initial wavefunction. The other solid curves correspond to the simulation of Eq. (\ref{eq:SchrodingerEquation}) and the dashed curves are the strong imaging limit of Eq. (\ref{eq:3level_rhoee}). (a) Simulation in the strong coupling regime corresponding to the experimental parameters in section \ref{sec:ho_cloud} ($\rm s_0$=0.022,$\ t_{\rm tr}=8\mu$s) and ($\rm s_0$=0.063,$\ t_{\rm tr}=12\mu$s) which are respectively in red and orange. Inset : full view of the wavefunction. (b) Simulation in the weak coupling regime corresponding to the experimental parameters in section \ref{sec:1Dlatt_cloud} ($\rm s_0$=0.02,$\ t_{\rm tr}=16\mu$s),$\ \omega=2\pi.130$ rad.kHz). The dotted line corresponds to the equilibrium solution derived from Eq. (\ref{eq:atom_number_weak_regime}). To highlight the difference between the two regimes, we show the wavefunction (red dashed line) if the strong imaging limit applies.}
\label{fig:numsim}
\end{center}
\end{figure}

{The diabatic and adiabatic behavior of the wavefunction expected respectively in the strong and weak imaging regime can be validated by numerical simulations of Eq. (\ref{eq:SchrodingerEquation}). For that purpose, using imaginary time method, we have derived the ground and few first eigenstate wavefunctions of $\hat{H}_0$. Starting from the ground state, we apply suddenly at time $t=0$ the loss term in Eq. (\ref{eq:SchrodingerEquation}) and numerically simulate the temporal evolution until $t_{\rm tr}$ at which the loss term is removed and the system measured. The results of such simulation are presented in Fig. \ref{fig:numsim} for the experimental parameters ($s_0$,$t_{\rm tr} (\mu$s)) corresponding to the data of the main text.\\
For the strong imaging regime (Figure \ref{fig:numsim} (a)), the initial state is a Gaussian state with size $\sigma_y=64\ \mu$m. The solid line is the simulation carried for parameters (0.022,8) and (0.063,12), which is slightly smoothed by the evolution but still very close to the strong imaging limit (dashed line) calculated from Eq. (\ref{eq:3level_rhoee}). The inset shows the entire wavefunction and the repumped region in red dashed line.\\
For the weak imaging regime (Figure \ref{fig:numsim} (b)), we start from the ground state of a 130 kHz harmonic oscillator frequency and repump with (0.02,12). The resulting simulated wavefunction (red solid line) strongly differs from the strong imaging limit (red dashed line). On the other hand, it corresponds very well to the adiabatic evolution (dotted blue line) which is calculated from Eq. (\ref{eq:atom_number_weak_regime}), and corresponds to a reduction of the initial wavefunction (solid blue line).} 


\section{Forces during excitation}\label{sec:forces_excitation}
{The imaging method involves the use of strong light shift gradients, possibly corresponding to non-negligible dipole forces during the time the atoms transit in the excited state. It might have two effects: a distortion of the ground state potential via a doubly dressed state (DDS) effect, and a strong acceleration of the atoms before being imaged by absorption imaging.\\
We compute the doubly dressed state potential corresponding to the average force that includes the ground and excited state potentials \cite{Bellouvet2018}:
\begin{equation}
    \begin{aligned} 
    U_{\text{DDS}}(x) =  \int_{-\infty}^{x} dy \left[ \rho_{5S}(\Delta)\frac{dU_{5S}}{dy}  +  \rho_{5P}(\Delta)\frac{dU_{5P}}{dy} \right]
    \label{eq:doubly_dressed_state_general}
    \end{aligned}
\end{equation}}

{Using $\rho_{5P}+\rho_{5S}=1$ and the detuning $\Delta=\Delta_{780}-(U_{5P}-U_{5S})$, Eq. (\ref{eq:doubly_dressed_state_general}) becomes:
\begin{equation}
    \begin{aligned} 
    U_{\text{DDS}}(x) = \int_{-\infty}^{x} dy \left[ \frac{dU_{5S}}{dy} - \rho_{5P}(\Delta)\frac{d\Delta}{dy} \right]
    \label{eq:doubly_dressed_state_general2}
    \end{aligned}
\end{equation}}

{Performing the integration with $\rho_{5P}(Y)=(s_0/2)/(1+s_0+Y^2)$ and setting the limits $U_{5S}(-\infty)=U_{5P}(-\infty)=0$, the doubly dressed state potential simplifies to:
\begin{equation}
\begin{aligned} 
U_{\text{DDS}}(x) & = U_{5S}(x) -\int_{2\Delta(-\infty)/\Gamma}^{2\Delta(x)/\Gamma} dY \rho_{5P}(Y) \\
& = U_{5S}(x) - \frac{s_0\Gamma}{4\sqrt{1+s_0}} \times \\ 
&\left( \atan(\frac{2\Delta(x)}{\Gamma\sqrt{1+s_0}}) - \atan(\frac{2\Delta_{780}}{\Gamma\sqrt{1+s_0}}) \right)
\label{eq:doubly_dressed_state_general3}
\end{aligned}
\end{equation}}

{We apply Eq. (\ref{eq:doubly_dressed_state_general3}) for our case at mid-fringe ($\Delta_{780}=U_{\text{5P,0}}/2$), aligned on a ground state lattice site, with a linearised detuning, and using $s_0\ll1$ to get: 
\begin{equation}
    \begin{aligned}
        U_{\text{DDS}}(x)  = U_{5S}(x)-\frac{s_0\Gamma}{4} \left(\atan(\frac{x}{X_0}) - \atan(\frac{U_{\text{5P,0}} }{\Gamma})  \right)
        \label{eq:doubly_dressed_state_general4}
    \end{aligned}
\end{equation}
The total Hamiltonian in the Schr\"{o}dinger equation now includes the potential given by Eq. (\ref{eq:doubly_dressed_state_general4}) and the kinetic energy such that $\hat{H}_{\text{tot}}=\hbar U_{\text{DDS}}(\hat{x})+\hat{H}_{\text{kinetic}}$. From this total Hamiltonian, we can extract an adiabaticity criterion where the effect of the strong light shifts does not excite motional states:
\begin{equation}
    \begin{aligned}
        \abs{\bra{\phi_n} \hat{H}_{\text{tot}} \ket{\phi_0} } \ll E_n-E_0 = n\hbar\omega_{\text{HO}}
        \label{eq:doubly_dressed_state_adiabaticity}
    \end{aligned}
\end{equation}
The sum of the 5S potential and the kinetic energy is diagonal in the state basis. Therefore, only the atan term in the potential creates off-diagonal couplings. As the excited state potential is odd, only odd couplings are non-zero and reads for $n=1$:
\begin{equation}
    \begin{aligned}
        \sqrt{\frac{\pi}{8}} \frac{\kappa_0}{c_{2'2}\omega_{\text{HO}}}e^{r_0^2}\text{erfc}(r_0) \ll 1.
    \label{eq:adiabaticity_DDS_limit}
    \end{aligned}
\end{equation}
In the strong localization limit ($r_0\ll1$), the term $e^{r_0^2}\text{erfc}(r_0)\approx1$. Therefore, it does not depend on the light shift amplitude, but only on the interaction strength set by $\kappa_0$. It is expected as $s_0$ determines the trap depth due to the excited state in the doubly dressed state potential (Eq. (\ref{eq:doubly_dressed_state_general4})). In our experimental settings where $r_0=0.24$, it reads $0.23\ll1$, which guarantees that the ground state potential is weakly perturbed by the strong light shift of the excited state. It is interesting to note that the effect of the strong gradient could in principle be neglected using a repumper laser resonant with the bottom of the excited state lattice where the derivative is zero.}



{Finally, let us estimate the impact of the acceleration on the atom number measurements after the transfer into $\ket{2}$ due to the force of the excited state. The force at mid-fringe is equal to $F_0=\pi U_{5P,0}/i_{1529}$. The atom velocity can be computed using the equation of motion $m\dot{v}=F_0$. After an average time in the excited state on the order of $1/\Gamma$, the atoms transferred into $\ket{2}$ have on average a velocity gain of $v=\pi U_{5P,0}/(m i_{1529}\Gamma)$. For an absorption imaging time $t_{im}=8$ \textmu s, the atoms moved by 0.088 \textmu m (resp. 0.4 \textmu m) in experiments with the thermal cloud (resp. the BEC) which is smaller than the size of a pixel (0.8 \textmu m) at the position of the atoms. Therefore, the absorption imaging is not impacted by this acceleration as the displacement of the atoms are smaller than the pixel size. Even for larger displacements, reliable atom number measurements can be done if the signal-to-noise per pixel is higher than 1, and if the Doppler effect along the imaging beam propagation direction is negligible (the acceleration due to the light shifts is in the transverse place).}


\section{Piezo displacement calibration}\label{sec:calibration_piezo}
\begin{figure}
\begin{center}
\includegraphics[scale=0.47]{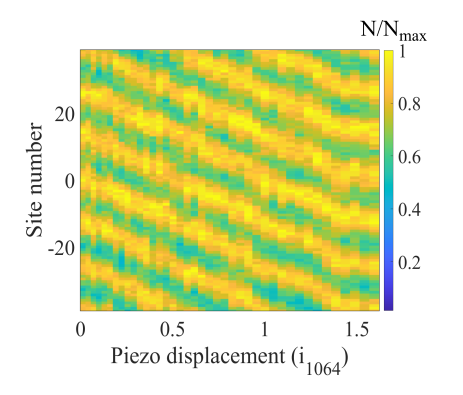}
\caption{Projection along the $x$-axis of normalized atom numbers after repumping atoms in the 1064 nm lattice with $\Delta_{780}=0$, $s_0=0.02$ and $t=8$ \textmu s in a modulation of $U_{\text{5P,0}}=16\Gamma$ as a function the camera axis ($y$-axis) and the relative phase $\Phi_0$ ($x$-axis) between the two lattices.}
\label{fig:piezo_displacement}
\end{center}
\end{figure}

The piezo stack controlling the position of the 1064 nm lattice is calibrated by measuring the displacement of the atomic fringes on a camera as a function of the piezo drive as shown on Fig. \ref{fig:piezo_displacement}. These fringes are obtained after loading the 1064 nm lattice and turning ON the 1529 nm lattice. The atoms are repumped  with $\Delta_{780}=0\Gamma$ and imaged by absorption imaging. The piezo displacement axis is ultimately calibrated by knowing that two consecutive fringes along the site number axis is equal to $13i_{1064}=6.9$ \textmu m. Along the piezo displacement axis, a displacement of exactly $i_{1064}$ gives the same fringe position due to the lattice periodicity. Note that the fringe visible around the site n$^{\circ}0$ at a displacement of $0.6i_{1064}$ corresponds to the site n$^{\circ}$-1 which is resonant. 

\section{Central position shifts}\label{sec:central_position_shifts}
\begin{figure}
\begin{center}
\includegraphics[scale=0.43]{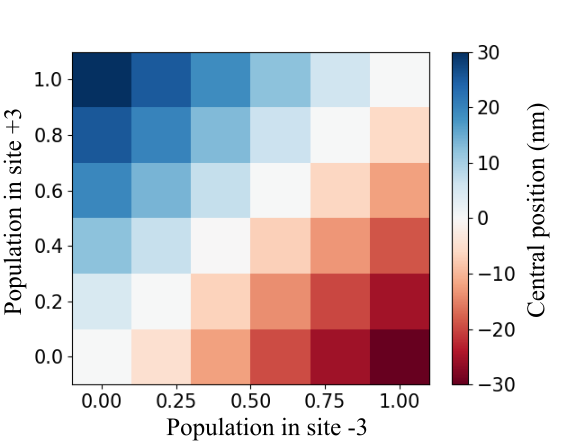}
\caption{Central position shift of the wavepacket as a function of relative initial populations in the sites $\pm3$.}
\label{fig:wavepacket_position_shift}
\end{center}
\end{figure}

We numerically compute the position shift of the imaged wavepacket by simulating the total repumped fraction for different relative populations in the sites $\pm3$ compared to the population in site 0. We scan numerically the relative phase between the lattices knowing the initial atomic density and the point spread function $\rho_{22}(x)$ from which we compute the repumped populations. Then, we fit the resulting wavepacket with a Gaussian function to extract the central position. We see on Fig. \ref{fig:wavepacket_position_shift} that for populations of approximately 0.6, the wavepacket shifts by $\pm20$ nm which corresponds to the experimental data. We checked that residual angles as described in Appendix \ref{sec:angle_sensitivity} do not shift significantly the position of the wavepacket.

\section{Width broadening by lattice misalignment}\label{sec:angle_sensitivity}
The atomic density in the ground state lattice site is characterized by the widths $\sigma_x\ll\sigma_{y,z}$ and $\sigma_{y}=\sigma_{z}$. Let us consider a 2D Gaussian atomic density in a rotated frame $(x',i')$ by an angle $\eta_i$ such that:
\begin{equation}
\begin{aligned}
n(x',i')=e^{-\frac{x'^2}{2\sigma_x^2}-\frac{i'^2}{2\sigma_i^2}}.
\label{eq:density_rotated}
\end{aligned}
\end{equation}

The rotated frame is expressed in the initial frame by:
\begin{equation}
\begin{aligned}
\begin{pmatrix}x'\\ i'\end{pmatrix} = \begin{pmatrix}\cos(\eta_i) & -\sin(\eta_i)\\\sin(\eta_i) & \cos(\eta_i)\end{pmatrix} \begin{pmatrix}x\\ i\end{pmatrix},
\label{eq:rotating_frame}
\end{aligned}
\end{equation}
where $i\in\{y,z\}$.

Integrating Eq. (\ref{eq:density_rotated}) over the direction $i$, using $\sigma_y\gg\sigma_x$ and considering the case of small angles $\eta_i\ll1$, one can compute the linear atomic density:
\begin{equation}
\begin{aligned}
n(x)=e^{-\frac{x^2}{2\tilde{\sigma}_x^2}},
\label{eq:density_rotated_integrated}
\end{aligned}
\end{equation}
where the effective width $\tilde{\sigma}_x$ is given in the main text. 

It is clear that a rotation about the $x$-axis does not lead to any broadening as the potential is symmetric.

\begin{figure}
\begin{center}
\includegraphics[scale=0.35]{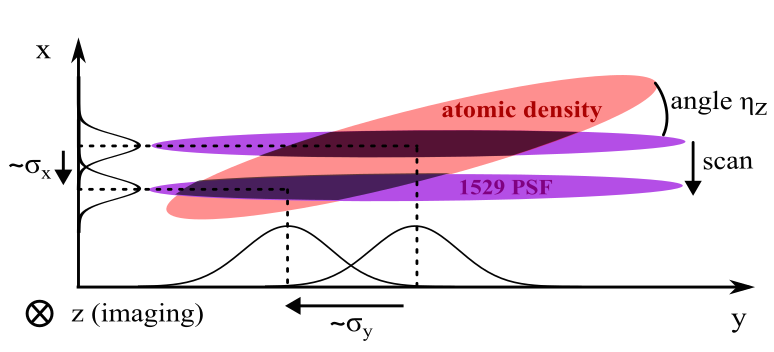}
\caption{Schematics of the projection position shift of the repumped atomic cloud in the case of a relative angle $\eta_z$ between an atomic wavepacket and a 1529 nm point spread function plane.}
\label{fig:relative_angle}
\end{center}
\end{figure}

In the case of a relative angle $\eta_z$ about the $z$-axis (Fig. \ref{fig:relative_angle}), the projection of the repumped cloud along $y$ is shifted on the order of $\sigma_y$ which corresponds to an angle of 0.3°. Experimentally, the cloud shifts by less than a micrometer so we believe that $\eta_z$ does not contribute to a broadening. Note that other datasets showed shifts on the order of a few micrometers. We could minimize these shifts by a better alignment of the counter-propagating 1529 nm laser beam.

The angle $\eta_y$ is difficult to estimate experimentally. On one hand, the $z$-direction is integrated by the imaging system so we do not have access to that spatial direction to look for a displacement of the projection. On the other hand, a small displacement on the order of $\sigma_x$ is difficult to measure along $x$.

As in Appendix \ref{sec:central_position_shifts}, we performed a complete simulation of the measured wavepacket but we also included $\eta_y$ as a free parameter. We use the initial population in the $\pm3$ sites that we found in Appendix \ref{sec:central_position_shifts}. For this population and using $\eta_y=0.3^{\circ}$, we numerically compute a width of 69 nm after the coarse cleaning step which matches with the experimental data of Fig. \ref{fig:counterPropasetupData}e in the main text. Also, we found that the expected width after the second cleaning step matches with the experimental width of $45\pm5$ nm. The effect of an angle is to globally shift the theoretical line towards larger widths.

Other sources of broadening could be consider in the case of perfect alignment. We ensured that the lattice laser frequencies do not drift significantly by frequency locking them on the same transfer cavity. Then, the initial phase at the atom position might also fluctuate over time due to room temperature changes. This would lead to an imperfect initial positioning of the site $0$ at the beginning of the first cleaning sequence. It would cause an asymmetry in the measured wavepacket in the second cleaning step as either the $-3$ site or $+3$ site would be closer or further from resonance. Finally, the last repumper pulse can be shortened to durations smaller than the trap frequency period to avoid dynamical effects. We tried to use shorter repumper durations and did not see a narrower wavepacket which suggests that the dominating width broadening was due a residual angle, as in such a case, the dynamics in the trap along $x$ would be on longer timescales than along $x'$.

\vspace{1cm}
\bibliography{library}

\end{document}